\documentclass{PoS}

\title{Beyond Standard Neutrino Theory}

\ShortTitle{Beyond Standard Neutrino Theory}

\author{\speaker{Toshihiko Ota}\\ 
        School of Physical Sciences and Engineering, Yachay Tech, Ecuador\\
        E-mail: \email{tota@yachaytech.edu.ec}}

\abstract{In this talk,
we will discuss phenomenology of new physics associated with neutrinos,
in particular,
non-standard neutrino interactions,
non-unitarity of the lepton mixing matrix,
and secret neutrino interactions mediated by a light field.
}

\FullConference{The 19th International Workshop on Neutrinos from Accelerators-NUFACT2017\\
		25-30 September, 2017\\
		Uppsala University, Uppsala, Sweden}

\usepackage{amsmath}	
\usepackage{mathrsfs}   
\usepackage{cite}

\begin{document}

\section{What is this talk about?}

The title given to me is ``Beyond Standard Neutrino Theory''
which can contain all the exotic features of neutrinos.
It is too broad to make it fit with a 25 minute talk.
Here, I would like to focus on the following three topics;
\#1. Non-standard neutrino interactions (NSI) parameterized with
four-Fermi interactions,
\#2. Non-unitarity of the Pontecorvo-Maki-Nakagawa-Sakata (PMNS) matrix,
and
\#3. Secret neutrino interactions,
where we will discuss the interactions between neutrinos and
invisible particles such as dark matter fields and also neutrinos.
This category mainly denotes a new force with neutrinos,
which is mediated by a field with a mass lighter than
the electroweak scale.

\section{Non-standard neutrino interactions}

\begin{figure}[t]
 \unitlength=1cm
  \begin{picture}(4,3.5)
   \put(-0.8,-0.2){\includegraphics[width=6.5cm]{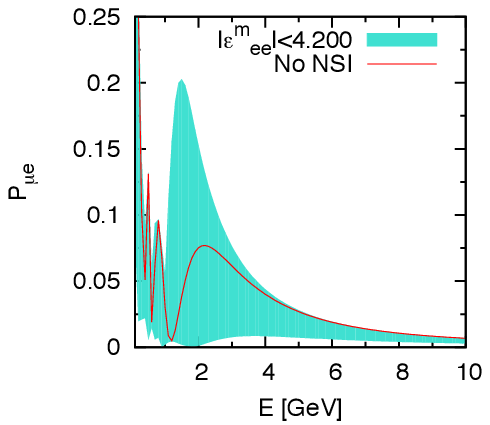}}
   \put(4.4,-0.2){\includegraphics[width=6.5cm]{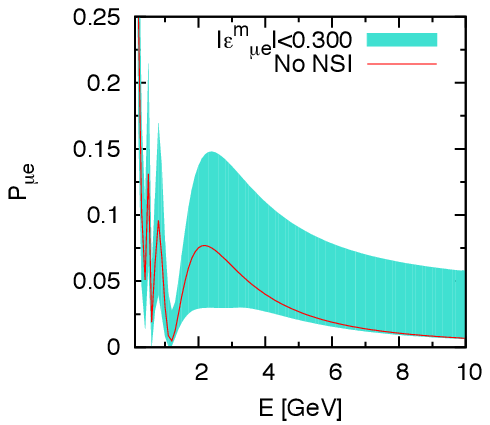}}
   \put(9.6,-0.2){\includegraphics[width=6.5cm]{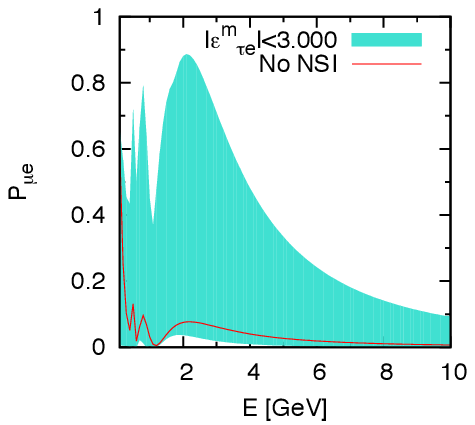}}
  \end{picture}
 \caption{Variations of the oscillation probability
 for $\nu_{\mu} \rightarrow \nu_{e}$ at the DUNE experiment
 with the change of the NSI parameters
 in their allowed ranges~\cite{Blennow:2016etl}.}
 \label{Fig:P-NSI-DUNE}
\end{figure}

NSIs are the four-Fermi interactions with neutrino,
which are expected to appear in general,
if the standard model is an effective theory of
fundamental theories realized at the high energy scales,
e.g.,
\begin{align}
 -
 \mathscr{L}_{\text{NC}}
 =&
 2\sqrt{2}
 G_{F}
 \epsilon_{\alpha \beta}^{m}
 [\overline{\nu}_{\alpha}
 \gamma^{\rho} {\rm P}_{L}
 \nu_{\beta}]
 [
 \overline{e} \gamma_{\rho} {\rm P}_{L} e
 ]+{\rm H.c.},
 \Longrightarrow (V_{\text{NSI}})_{\alpha \beta} =
 \sqrt{2} G_{F} N_{e}
 \epsilon^{m}_{\alpha \beta},
 \\
 -
 \mathscr{L}_{\text{CC}}
 =&
 2\sqrt{2}
 G_{F}
 \epsilon_{\mu \alpha}^{s}
 [\overline{\nu}_{\alpha}
 \gamma^{\rho} {\rm P}_{L}
 \mu]
 [
 \overline{d} \gamma_{\rho} {\rm P}_{L} u
 ]+ {\rm H.c.},
 \Longrightarrow
 |\nu_{\mu}^{s} \rangle =
 |\nu_{\mu} \rangle
 +
 \epsilon^{s}_{\mu \alpha} |\nu_{\alpha} \rangle
 \text{ in $\pi^{+}$ decays}.
\end{align}
The neutral current (NC) type NSI with two neutrinos affects
the neutrino propagation Hamiltonian as an additional matter effect
potential term $V_{\text{NSI}}$.
If we have a charged current (CC) type NSI with a neutrino
and a charged lepton, it modifies the initial 
and the final states in neutrino oscillation amplitudes.
There is still a plenty of room left for them.
Particularly, the NC type NSIs are still phenomenologically
allowed to be as large as the order of 0.1
or even the order of 1.
For the current upper bounds of the NSI parameters,
see e.g., Ref.~\cite{Choubey:2015xha}.
As we will see later, it is not so easy to have NSIs with the size
of unity in a theoretical point of view. However, we should not
exclude the possibilities allowed phenomenologically.
In Fig.~\ref{Fig:P-NSI-DUNE} taken from Ref.~\cite{Blennow:2016etl},
one can find how much the oscillation probability for $\nu_{\mu}$
to $\nu_{e}$ at the DUNE experiment can be modified
by the variations of the NSIs in their allowed ranges.
Here I would like to overview the phenomenological studies on NSIs,
categorizing them into two groups;
One is NSIs as noise in the determination of the standard oscillation
parameters and the other is NSIs as a signal of new physics.
%

\subsection{NSIs as noise}

\begin{figure}[t]
 \unitlength=1cm
 \begin{picture}(16,3.5)
  \put(-0.1,0.3){\includegraphics[width=5.5cm]{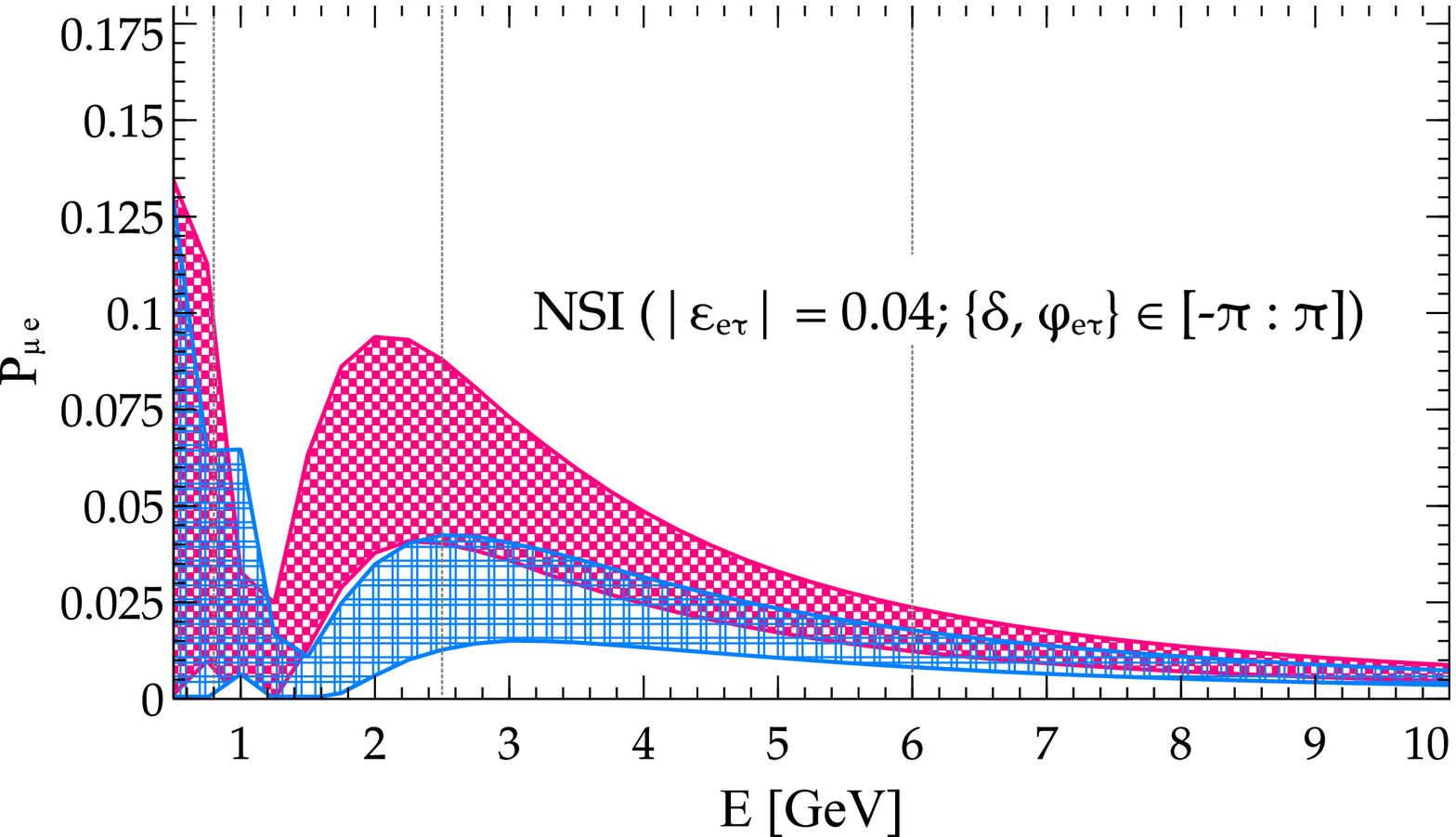}}
  \put(5.5,-0.1){\includegraphics[width=5.8cm]{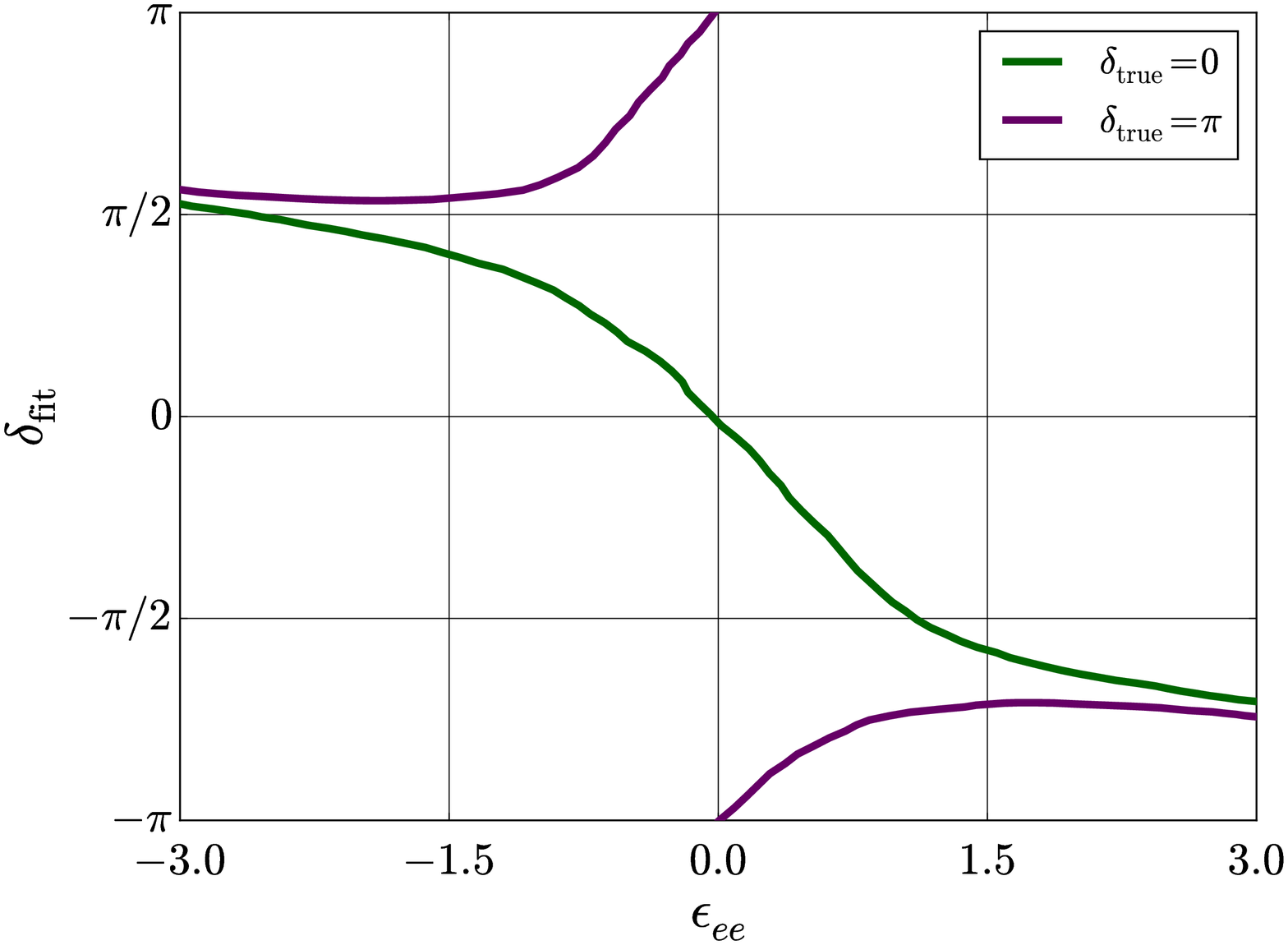}}
  \put(11,0){\includegraphics[width=4cm]{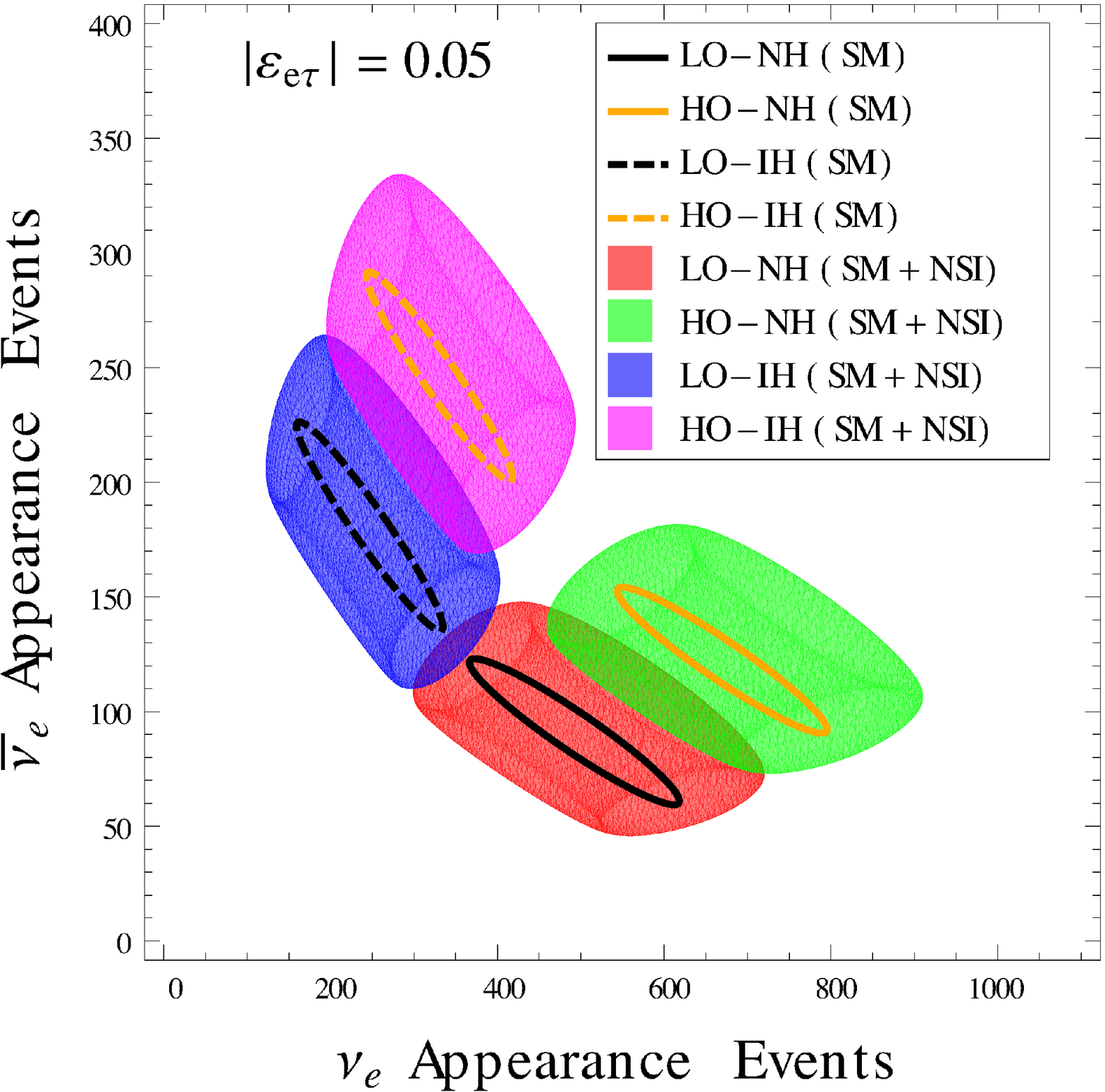}}
 \end{picture}
 \caption{NSIs as noise; NSIs disturb the determination of
 the standard oscillation parameters at the DUNE experiment.
 [Left] With $|\epsilon^{m}_{e\tau}|=0.04$,
 the oscillation probability of NH (red)
 overlaps with that of IH (blue)~\cite{Masud:2016gcl}.
 [Middle] The best-fit value $\delta_{\text{fit}}$
 of the CP violating phase is
 shifted by the introduction of
 $\epsilon^{m}_{ee}$~\cite{deGouvea:2016pom}.
 [Right]
 With $|\epsilon^{m}_{e\tau}|=0.05$,
 the ellipses for the lower octant (LO) cases in the bi-probability plane 
 is no more clearly separated from that for
 the higher octant (HO) cases~\cite{Agarwalla:2016fkh}.
 }
 \label{Fig:NSI-noise}
\end{figure}

NSIs disturb the determination of the standard oscillation parameters.
Here I would like to
discuss the impact of the noise made by the NSIs
at the DUNE experiment.
%
The normal hierarchy (NH) case in the standard oscillation scenario
is clearly separated from the inverted hierarchy (IH) case at
the beam peak (2-3 GeV) of DUNE.
However, these two cases start overlapping each other,
if NSIs are introduced
with the size of the order of $10^{-2}$, as shown in the left plot of
Fig.~\ref{Fig:NSI-noise}~\cite{Masud:2016gcl}, 
and consequently,
the sensitivity to the mass hierarchy
is reduced~\cite{Masud:2016gcl,Dutta:2016czj,Deepthi:2016erc}.
The NSIs also confuse the fit value of the CP violating phase
$\delta$~\cite{Masud:2015xva,Masud:2016bvp,Rout:2017udo,Miranda:2016wdr,deGouvea:2016pom,Ge:2016dlx,Dutta:2016vcc}.
The middle plot in Fig.~\ref{Fig:NSI-noise}~\cite{deGouvea:2016pom}
shows
that
the best-fit value $\delta_{\text{fit}}$ suggested by
the DUNE experiment is shifted from the true value
$\delta_{\text{true}}$
by the introduction of $\epsilon^{m}_{ee}$.
A serious problem here is,
since the fit is not bad (the $\chi^{2}$ is reasonably small),
the result looks consistent with the standard oscillation scenario,
and therefore,
we cannot tell if we have exotic effects or not.
NSIs also affect the determination of the octant of the 2-3 mixing
angle~\cite{Agarwalla:2016fkh,Dutta:2016eks,Das:2017fcz};
See the right plot in Fig.~\ref{Fig:NSI-noise}~\cite{Agarwalla:2016fkh}.
The NSIs can be a serious noise of the parameter determination
at the forthcoming long baseline experiments
--- How can we reduce this noise?

\begin{figure}[t]
 \unitlength=1cm
  \begin{picture}(15,4)
   \put(0.2,-0.3){\includegraphics[width=6.8cm]{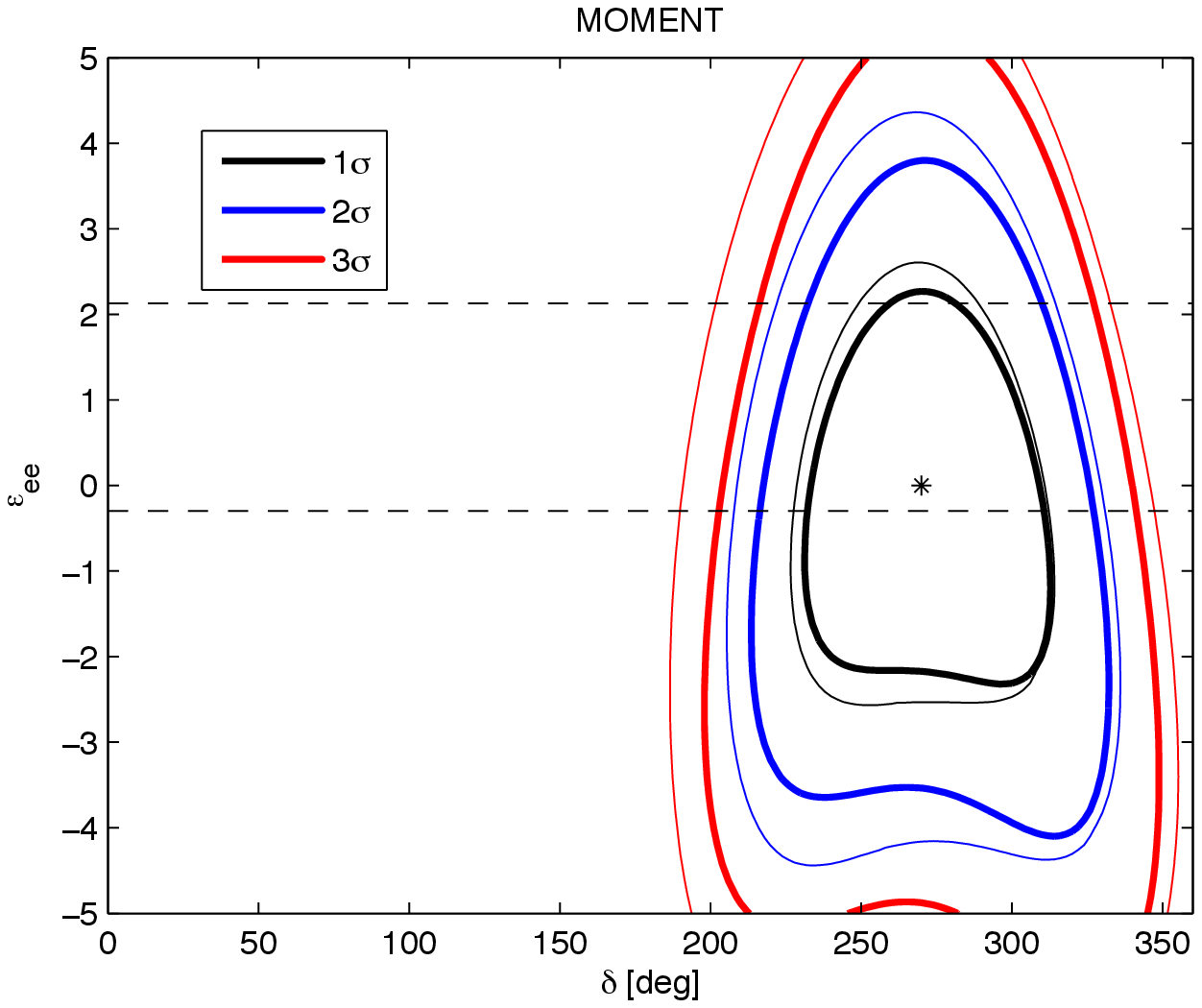}}
   \put(8,-0.1){\includegraphics[width=6.8cm]{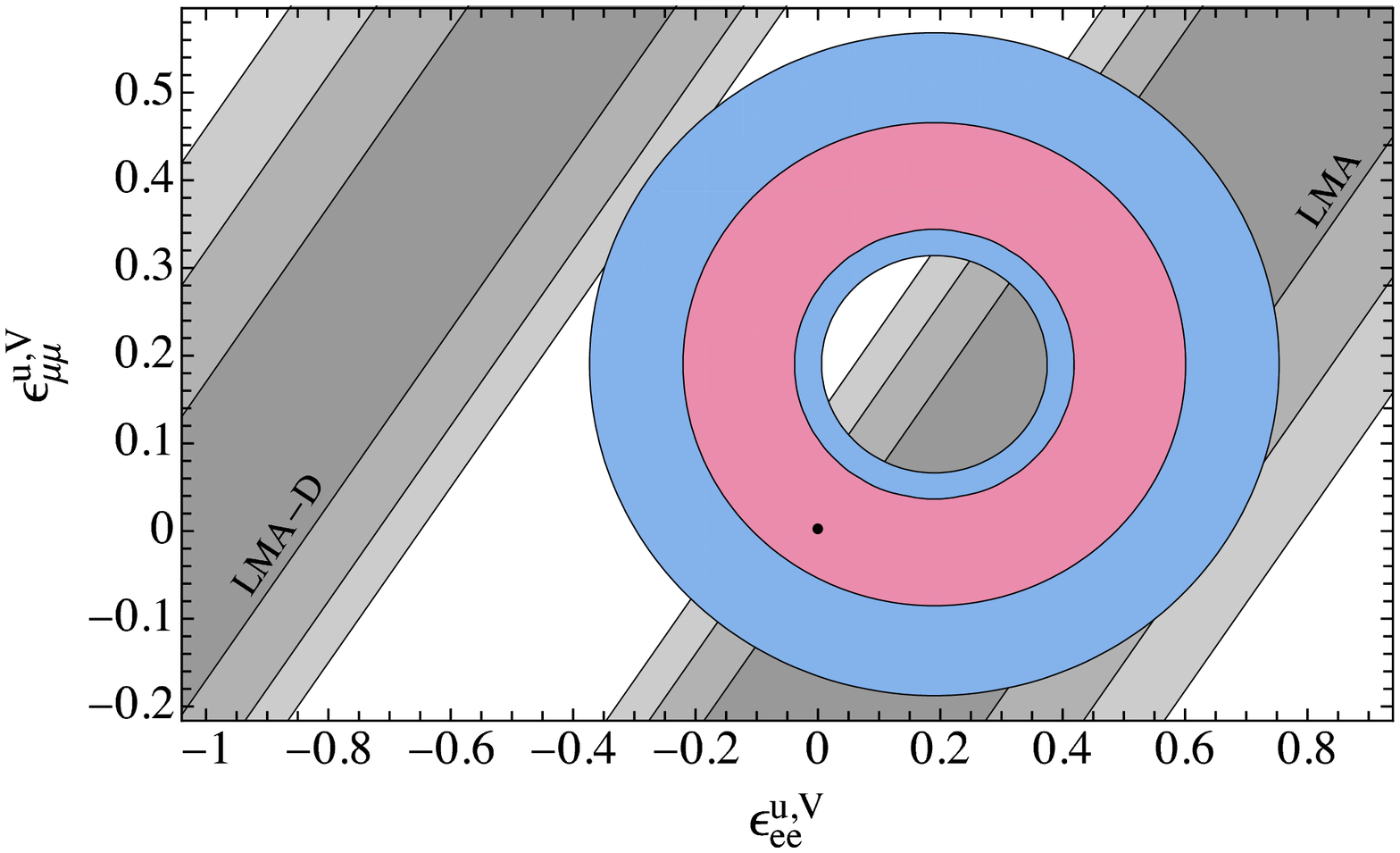}}
  \end{picture}
 \caption{To reduce the noise from NSIs in the
 determination of the standard oscillation parameters.
 [Left] Combination with an experiment with a shorter baseline;
 The MOMENT experiment with the baseline of 150 km
 can determine $\delta$ regardlessly of the matter NSIs~\cite{Bakhti:2016prn}.
 [Right] Constraints from non-oscillation experiments;
 The recent COHERENT result excludes the NSIs
 for the dark solar solution in the case where the NSIs are
 mediated by a heavy field~\cite{Coloma:2017ncl}.}
 \label{Fig:NSI-noise-reduction}
\end{figure}

Here I would like to discuss three possibilities.
The first thing we can do is
to combine the result of T2(H)K which has shorter baseline
than DUNE has. Thanks to that,
T2(H)K is less affected by the matter effect and the matter
related effects including the NC type NSIs.
The authors of Ref.~\cite{Ge:2016dlx}
shows that
the best-fit value of $\delta$ at T2K stays at the true value point
even with the $\epsilon^{m}_{ee}$ with the size of $\sim 1$,
although the $\chi^{2}$ for the fit of $\delta$ is damaged a bit.
This study suggests that T2(H)K can determine $\delta$ regardlessly
of the existence of the matter NSIs.
After the determination of $\delta$ at T2(H)K,
the determination of mass hierarchy
(and discrimination of non-standard effects)
at DUNE becomes easier ---
DUNE and T2HK are a nice synergetic combination.
We can take a step forward to this direction;
Having a new experiment with a shorter baseline
and a good profile beam.
As shown in the left plot in Fig.~\ref{Fig:NSI-noise-reduction}
taken from Ref.~\cite{Bakhti:2016prn},
the fit value of $\delta$ (horizontal axis)
at the MOMENT experiment with a baseline of
150 km does not depend on the value of the NSI $\epsilon^{m}_{ee}$
at all.
There are some plans of muon-based neutrino sources
---
MOMENT (muon decay in flight)~\cite{Cao:2014bea},
DAEdALUS/IsoDAR (muon decay at rest)~\cite{Aberle:2013ssa},
and NuSTORM (muon storage ring)~\cite{Kyberd:2012iz,Adey:2013pio}.
The neutrino beam spectra from
those sources are very precisely known, and the intensity of those
beams is expected to be high.
A high-intensity and relatively low-energy conventional neutrino beam
at the European Spallation Source (ESS) is also planed~\cite{Baussan:2012cw}.
Such short-baseline, low-energy, and high-intensity oscillation
experiments help reduce the noise from matter NSIs in the determination
of the standard oscillation parameters.

The second way to reduce the noise is the use of the results of
non-oscillation experiments.
In fact, the current experimental bounds to the NSIs
are mainly placed by non-oscillation experiments.
Recently, the COHERENT experiment has measured the 
coherent elastic neutrino-nucleon scattering (CE$\nu$NS)
process for the first time in the history~\cite{Akimov:2017ade} and
shows that the cross section is consistent with the standard model
prediction.
The result suggests that 
the NSI parameters
for the Dark solar solution~\cite{Miranda:2004nb}
is now excluded at 3 sigma (cf. the right plot
in Fig.~\ref{Fig:NSI-noise-reduction}~\cite{Coloma:2017ncl})
if the mass of the field that mediates the NSIs is taken to
be heavier than dozens of MeV~\cite{Liao:2017uzy}.
This example tells us that
the improvement of non-oscillation
neutrino experiments is crucially important
to exclude the possible disturbance from large NSIs.

Finally, I would like to remark the importance of tracing out
the oscillation curves in a wide range of $L/E$.
The distortion in the oscillation curves,
which is caused by NSIs,
has a characteristic $L/E$ dependence.
In the high energy limit with a fixed $L$,
all the standard oscillation
terms drop quickly as $1/E^{2}$ or $1/E^{3}$.
The NSI terms
of $\epsilon^{m}_{e\mu}$ in $\nu_{\mu} \rightarrow \nu_{e}$,
can stay in the high energy limit,
because they depend on $1/E$, and consequently,
it is easy to extract the information of this NSI effect
from the oscillation probability at the high energy region.
We can separate the NSI effects from the standard oscillation effects
and also can separate the different NSI effects
by observing the oscillation curves in a wide energy range.
Under the standard oscillation assumption, a focused-energy
(narrow-band) beam
is convenient to determine
the oscillation parameters.
However, scanning the oscillation probabilities over a wide energy
region might be also important to discriminate the non-standard
effect and establish the standard oscillation
scenario (or discover the exotic effects).

\subsection{NSIs as signal}

\begin{figure}[t]
 \unitlength=1cm
  \begin{picture}(15,2.5)
   \put(-0.2,0){\includegraphics[width=7.5cm]{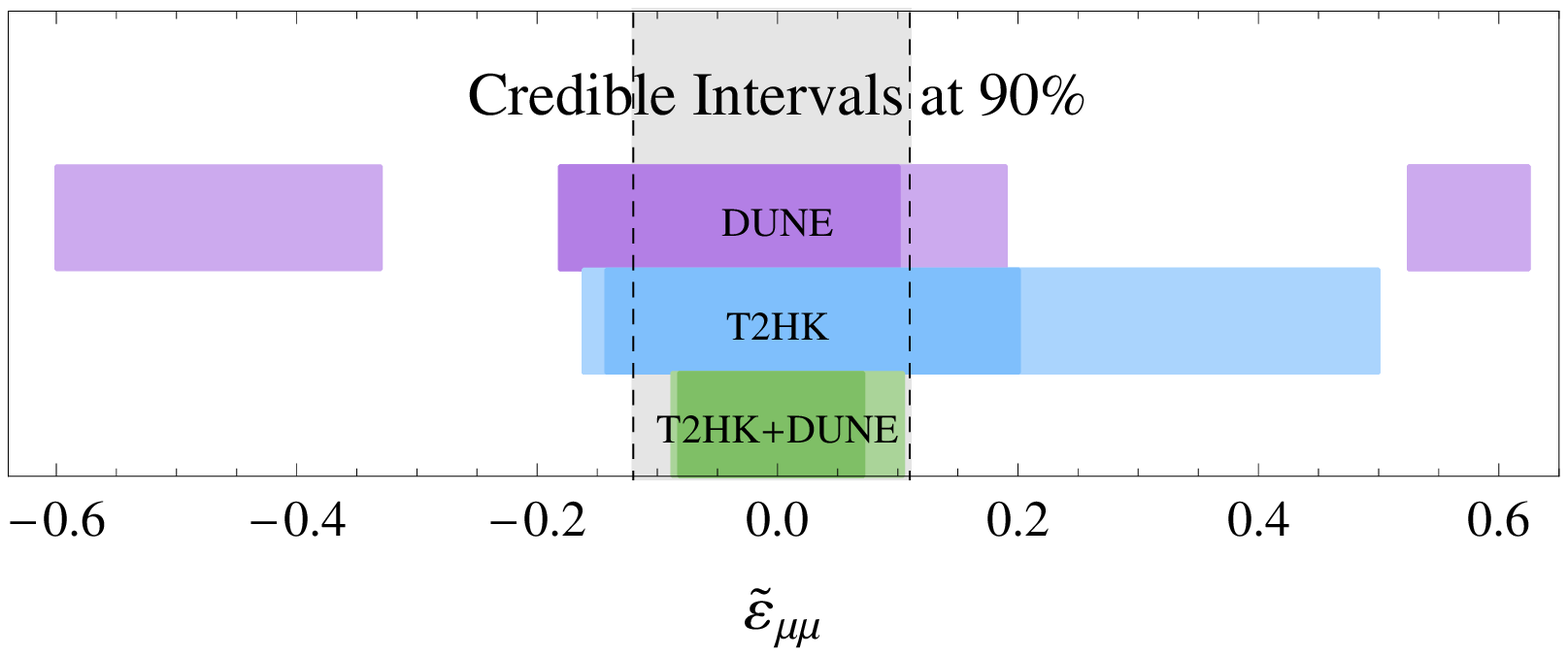}}
   \put(7.8,0){\includegraphics[width=7.5cm]{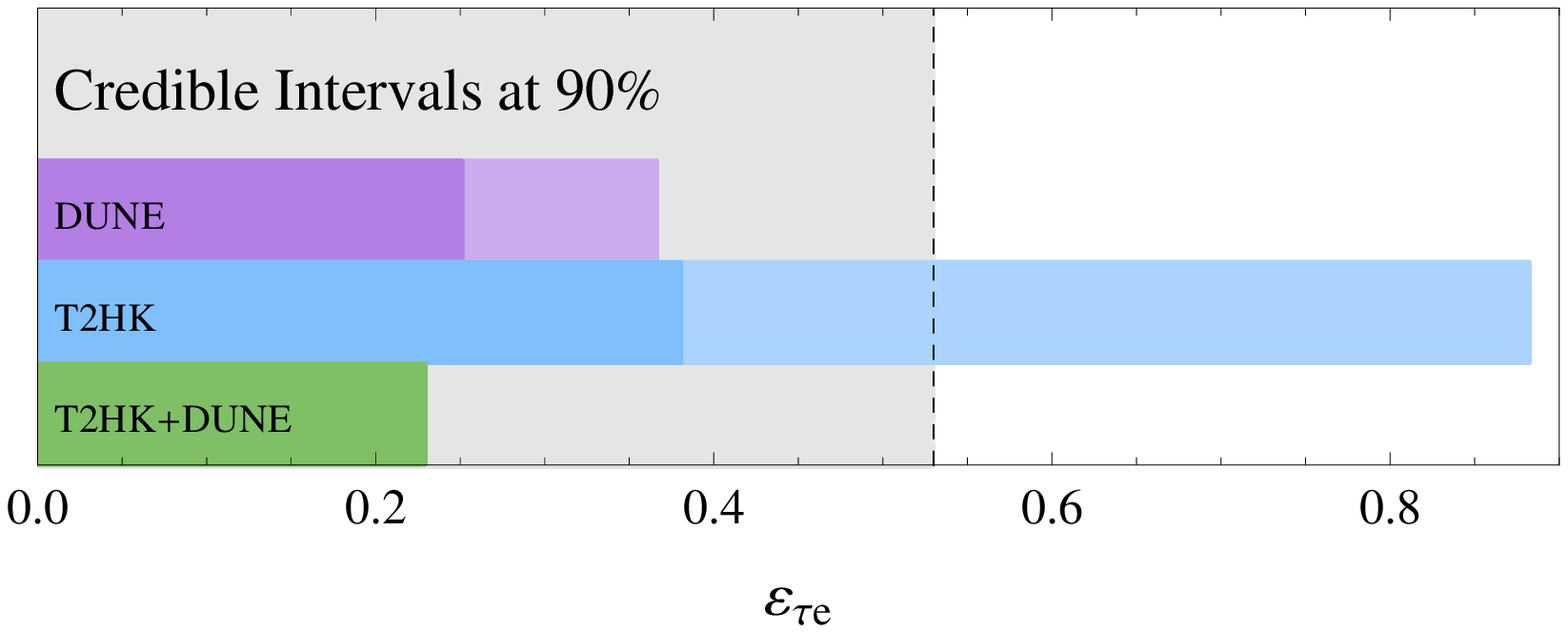}}
  \end{picture}
 \caption{NSIs as signal;
 Sensitivities of DUNE, T2HK, and their combination to
 $\tilde{\epsilon}^{m}_{\mu \mu} \equiv \epsilon^{m}_{\mu \mu} -
 \epsilon^{m}_{\tau \tau}$ and $\epsilon^{m}_{e\tau}$~\cite{Coloma:2015kiu}.}
\label{Fig:NSI-sensitivities} 
\end{figure}

There are also many studies on the sensitivities to NSIs at future
oscillation experiments ---
Not only at accelerator based experiments but also atmospheric
neutrino experiments.
Now, the IceCube starts participating in this game~\cite{Salvado:2016uqu,Aartsen:2017xtt}.
At Fig.~\ref{Fig:NSI-sensitivities},
I would like to introduce a few plot
for the sensitivities at DUNE and T2HK,
which is taken from Ref.~\cite{Coloma:2015kiu}.
With the combination of these two experiments,
we can explore the NSIs with the size of 0.1-0.05.
For the sensitivities at the other experiments,
see, e.g., Refs.~\cite{Fukasawa:2016nwn,Kelly:2017kch}
(T2HK and atmospheric neutrino at HK),
Ref.~\cite{Choubey:2015xha} (INO)
and
Ref.~\cite{Coelho:2017cwp} (ORCA-KM3NeT).
The sensitivities to the CC type NSIs
at future high intensity facilities
are studied in 
Refs.~\cite{Tang:2017qen} (MOMENT)
and \cite{Blennow:2015nxa} (ESS$\nu$SB).

\subsection{Model building for NSIs}

Here I would like to discuss briefly
some theoretical aspects of NSIs.
Large NSIs such as order of unity are still phenomenologically allowed
--- Even some of the small tensions in oscillation experiments suggest
large NSIs~\cite{Liao:2017awz,Liao:2016bgf,Fukasawa:2016gvm}.
Of course, we should pay attention to the possibilities
allowed phenomenologically.
However, it is not so easy to accommodate with large NSIs.
Naively thinking, a mass dimension-six four-Fermi interactions
consists of two fundamental interactions
mediated by a field such as $Z'$, a leptoquark, etc.
Setting the couplings of the fundamental interactions
to be unity and the mass of the mediator to be TeV,
we have the NSI with the size of the order of $10^{-2}$
for the effective coupling $\epsilon$,
which is a kind of a natural size of NSIs we can expect.
However, if there is a TeV field with such a large coupling
with charged fermions, we should have
already found it at the LHC --- This naive estimation tells us
that the NSIs should be smaller than $\mathcal{O}(10^{-2})$.
In addition, the NSIs originated from this type of dimension-six
operators appear always with the $SU(2)$ counter part
with charged leptons.
We have to take account of the constraints from the charged lepton
counter process.
The exhaustive studies on the dimension-six NSIs were already done in
1990s~\cite{Grossman:1995wx,Bergmann:1998ft,Bergmann:1999rz,Bergmann:1999pk},
and it turned out that the charged lepton processes constrain
the NSIs at the level of $10^{-3}$ in general.
To avoid the constraints from the charged lepton processes and enjoy
the upper bounds of NSIs,
the use of dimension-eight operators
$(\bar{L}L)(\bar{f}f)(H^{\dagger}H)$
with four fermions
and two Higgs fields was proposed~\cite{Berezhiani:2001rs,Davidson:2003ha}.
The dimension-eight operators do not contain the charged lepton
counter part, because the $SU(2)$ on the operators is violated
with the Higgs fields.
However, they are not free from everything.
Here I would like to introduce two arguments
on the constraints to the dimension-eight NSIs.
First, we have to think about the origin of the effective operators.
%
If the NSI is so large as order unity, it should be mediated
by tree-level diagrams.
The decomposition of the operator to the tree-level diagrams
leads to either dimension-six NSIs $(\bar{\nu}\nu)(\bar{f}f)$
or dimension-six non-unitary operators
$(\bar{\nu}\partial\cdot\gamma \nu)(H^{\dagger}H)$
or both~\cite{Antusch:2008tz,Gavela:2008ra}.
We can still think of the situation where those dimension-six operators
are, somehow, cancelled and only the 
counter-part-free dimension-eight-origin
NSI survives.
However, the second argument comes.
With the dimension-eight operators
such as $(\bar{L}L)(\bar{f}f)(H^{\dagger}H)$,
one can close the outer lines and draw one-loop dimension-six
operator $(\bar{L}L)(\bar{f}f)$, and 
the loop contribution
is quadratically divergent~\cite{Biggio:2009kv,Biggio:2009nt}.
To regularize this divergence, we need to introduce
the dimension-six operator $(\bar{L}L)(\bar{f}f)$
as the counter term.
Consequently, we have to have the dimension-six operator in the effective
Lagrangian. This means, we are essentially working with
dimension-six  NSI models.
It may be not impossible to solve all these dimension-six mess
by adding the terms and tuning the couplings,
but as we have seen, it is not so easy to obtain NSIs
with the size of their upper bound.

Recently, an alternative way has been discussed to obtain large NSIs,
which is NSIs mediated with light mediators~\cite{Farzan:2015doa,Farzan:2015hkd,Farzan:2016wym,Machado:2016fwi,Babu:2017olk}.
One can obtain the NSIs with the size of $10^{-2}$
with a $\sim$10 MeV mediator with a faint interaction with
the couplings of $\sim 10^{-5}$,
instead of the TeV mediator with order-one couplings.
A new force mediated by a light field is a kind of trend now,
and
there are many discussions recently in various contexts,
such as
cosmic neutrino spectrum,
muon anomalous magnetic moment,
cosmology (e.g., Refs~\cite{Farzan:2015pca,Huang:2017egl}),
lepton flavour non-universality in $B$-physics
(e.g., Ref.~\cite{Datta:2017pfz}),
etc.
Neutrino interactions with a light mediator are often
labeled as secret neutrino interactions.
In the last section,
I would like to discuss a scenario of the light leptonic force
related to the cosmic neutrino spectrum and muon $g-2$.

\section{Non-unitarity}

\begin{figure}[t]
\unitlength=1cm
\begin{picture}(15,10)
 \put(0,0){\includegraphics[width=15cm]{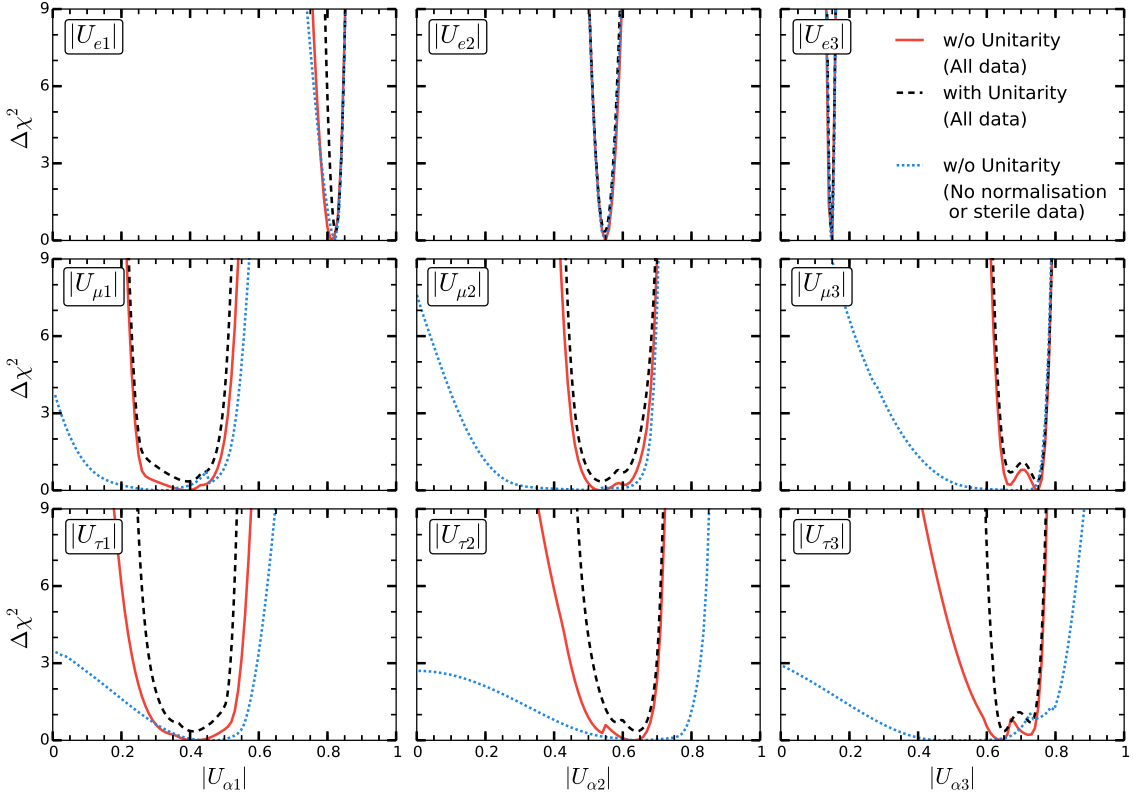}}
\end{picture}
 \caption{Fit of the elements of the PMNS matrix
 without the assumption of the unitarity (red)~\cite{Parke:2015goa}.}
 \label{Fig:Nonuni}
\end{figure}

Let us turn the topic to the unitarity violation.
Although the unitarity is a fundamental feature which is
related to the conservation of the probability,
the unitarity in the system of active neutrinos
can be violated in general.
The 3-by-3 PMNS matrix may be a part of the full lepton mixing matrix
which is unitary.
The extra flavours/generations can or cannot participate in
oscillation, depending on their masses
(e.g., Ref.~\cite{Fong:2016yyh}).
If the extra generations are light enough to participate in
oscillations,
we can expect the oscillation signals driven by the extra
$\Delta m^{2}$s.
If the extra generations are heavy and cannot participate in the
neutrino propagation process,
we do not have the extra oscillation signals
but the mixing matrix appears in the oscillation probabilities
becomes non-unitary~\cite{Langacker:1988ur,Antusch:2006vwa}
---
The scenario that neutrinos are mixing with heavy neutral fields
is called Minimal Unitarity Violation (MUV)~\cite{Antusch:2006vwa}.
If neutrinos decay into invisible debris, the Hamiltonian that
describes the time evolution of the neutrino system is
no more hermitian~\cite{Berryman:2014yoa,Coloma:2017zpg,Choubey:2017dyu}.
In such a case, the mixing matrix is not unitary in general.

Our current knowledge on the elements of the PMNS matrix
is nicely summarized in Fig.~\ref{Fig:Nonuni},
which is taken from Ref.~\cite{Parke:2015goa}
--- The authors fit the elements
without assuming the unitarity relations of the PMNS matrix.
Since the current oscillation results can be explained with
the 3-by-3 unitary PMNS matrix,
the best-fit values of all the elements
are consistent with the unitary case.
Here you can see that the normalization of the oscillation
probabilities set by no-oscillation results in short baseline
experiments are important to constrain the elements, in particular,
in the $\mu$ and the $\tau$ rows.

In the MUV regime,
the non-unitary PMNS matrix $N$ can be parameterized with
a hermitian matrix
$\eta_{\alpha\beta}$ ($\alpha,\beta\in\{e,\mu,\tau\}$)
as
\begin{align}
 (N_{\text{PMNS}})_{\alpha i}
 =
 (\delta_{\alpha \beta} + \eta_{\alpha \beta}) \mathcal{U}_{\beta i} 
\end{align}
where $\mathcal{U}$ is the unitary part.\footnote{%
There is another popular way to parameterize the deviation from the
unitarity in
the MUV~\cite{Escrihuela:2015wra,Miranda:2016wdr,Ge:2016xya,Escrihuela:2016ube}.
For the relation between two parameterizations
and the current and future bounds in both the parameterization schemes,
see Ref.~\cite{Blennow:2016jkn}.}
The elements of $\eta_{\alpha \beta}$ are constrained from
the mass of the $W$ boson, the Weinberg angle, muon decay rate,
lepton-flavour-violating processes
etc. at the level of $10^{-3}$ to 0.05~\cite{Antusch:2014woa,Antusch:2016brq,Fernandez-Martinez:2016lgt,Blennow:2016jkn,Escrihuela:2016ube}.
The non-unitarity $\eta$ modifies oscillation probabilities as
a combination of the NSIs~\cite{Meloni:2009cg,Blennow:2016jkn},
and the size of the $\epsilon$ parameters
induced by the non-unitarity effect corresponds in general
to the order of $\eta$.
Therefore, the $\eta$ with the size of 0.05 can be a noise
in the parameter determinations at the forthcoming experiments.
The constraints to the non-unitarity in the MUV scenario
at the future oscillation experiments are discussed
in e.g. Refs.~\cite{Blennow:2016jkn,Tang:2017khg}.

\subsection{Unitarity triangle}

We have seen that the current oscillation results
are consistent with the unitary PMNS matrix.
Is there a way to check the unitarity through the oscillations?
---
This question was addressed
in Refs.~\cite{Sato:2000wv,Farzan:2002ct,Parke:2015goa,Pas:2016qbg}.
In principle,
the unitarity of the PMNS matrix
can be checked by using one oscillation channel.
With the unitary PMNS matrix $U_{\alpha i}$,
the oscillation probability in vacuum for $\nu_{\mu} \rightarrow \nu_{e}$
consists of the following four terms at the orders of $|U_{e3}|^{2}$,
$\Delta m_{21}^{2} |U_{e3}|$, and $(\Delta m_{21}^{2})^{2}$;
\begin{align}
 P_{\nu_{\mu} \rightarrow \nu_{e}}^{\text{Vac}}
 =&
 \overbrace{4 |U_{\mu 3} U_{e3}^{*}|^{2}}^{\text{coeff. }A}
 \sin^{2} \frac{\Delta m_{31}^{2} L}{4E}
 +
 \overbrace{4 {\rm Re}[U_{\mu 2} U_{e2}^{*} U_{\mu 3}^{*} U_{e3}]}^{
 \text{coeff. }B}
 \left[
 \frac{\Delta m_{21}^{2} L}{4E}
 \right]
 \sin \frac{\Delta m_{31}^{2} L}{2E}
 \nonumber
 \\
 &-
 \underbrace{
 8 {\rm Im}[U_{\mu 2} U_{e2}^{*} U_{\mu 3}^{*} U_{e3}]
 }_{\text{coeff. }C}
 \left[
 \frac{\Delta m_{21}^{2} L}{4E}
 \right]
 \sin^{2} \frac{\Delta m_{31}^{2} L}{4E}
 +
 \underbrace{
 4 |U_{\mu 2} U_{e2}^{*}|^{2}
 }_{\text{coeff. }D}
 \left[
 \frac{\Delta m_{21}^{2} L}{4E}
 \right]^{2}.
\end{align}
Since all the terms have different energy dependencies,
one can separate them by tracing out the oscillation probability
in a wide energy range
and can know the coefficients of these four terms independently.
Combining the information of the coefficients $A$, $D$,
(lengths of the sides), and $B$ (the angle between the sides),
one can construct a unitarity triangle.
The coefficient $C$ is the Jarlskog invariant which is directly
proportional to the area of
the triangle~\cite{Jarlskog:1985ht,Jarlskog:1985cw}.
In short, we can check the unitarity by comparing the information
brought by $A$, $B$, and $D$ 
and the information of
$C$~\cite{Sato:2000wv,Pas:2016qbg}.
This method is spoiled at the high energy region,
because the $A$, the $B$, and the $D$ terms
behave in the same way as $\propto 1/E^{2}$
and they cannot be separated any more.
Therefore, to carry out this unitarity check,
we need to measure the oscillation probability with high precision
in the relatively low-energy region.
It is tough in the experimental sense, but in principle,
we can do.

\section{Secret neutrino interactions}

\begin{figure}[t]
 \unitlength=1cm
  \begin{picture}(15,4)
   \put(0.5,-0.4){\includegraphics[width=5.8cm]{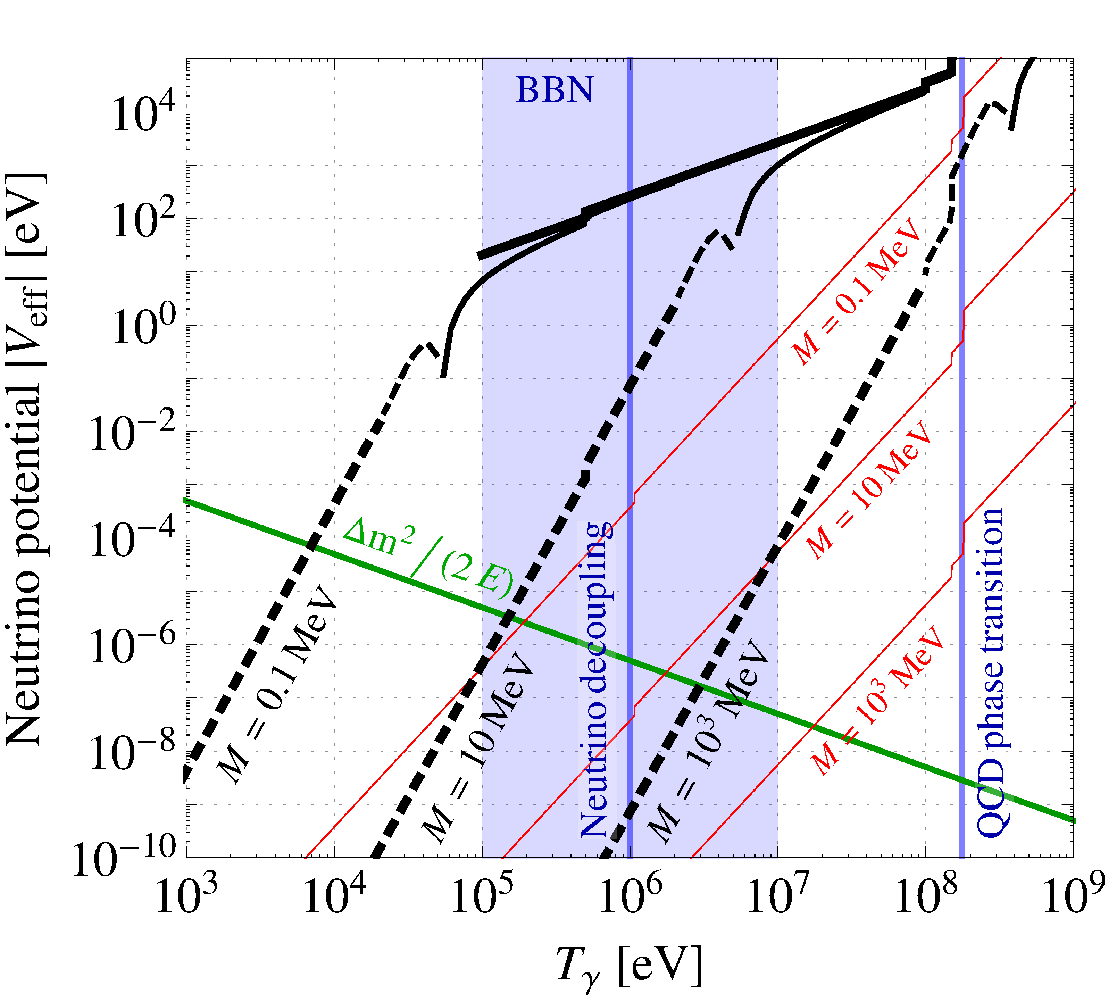}}
   \put(8.5,0){\includegraphics[width=6.5cm]{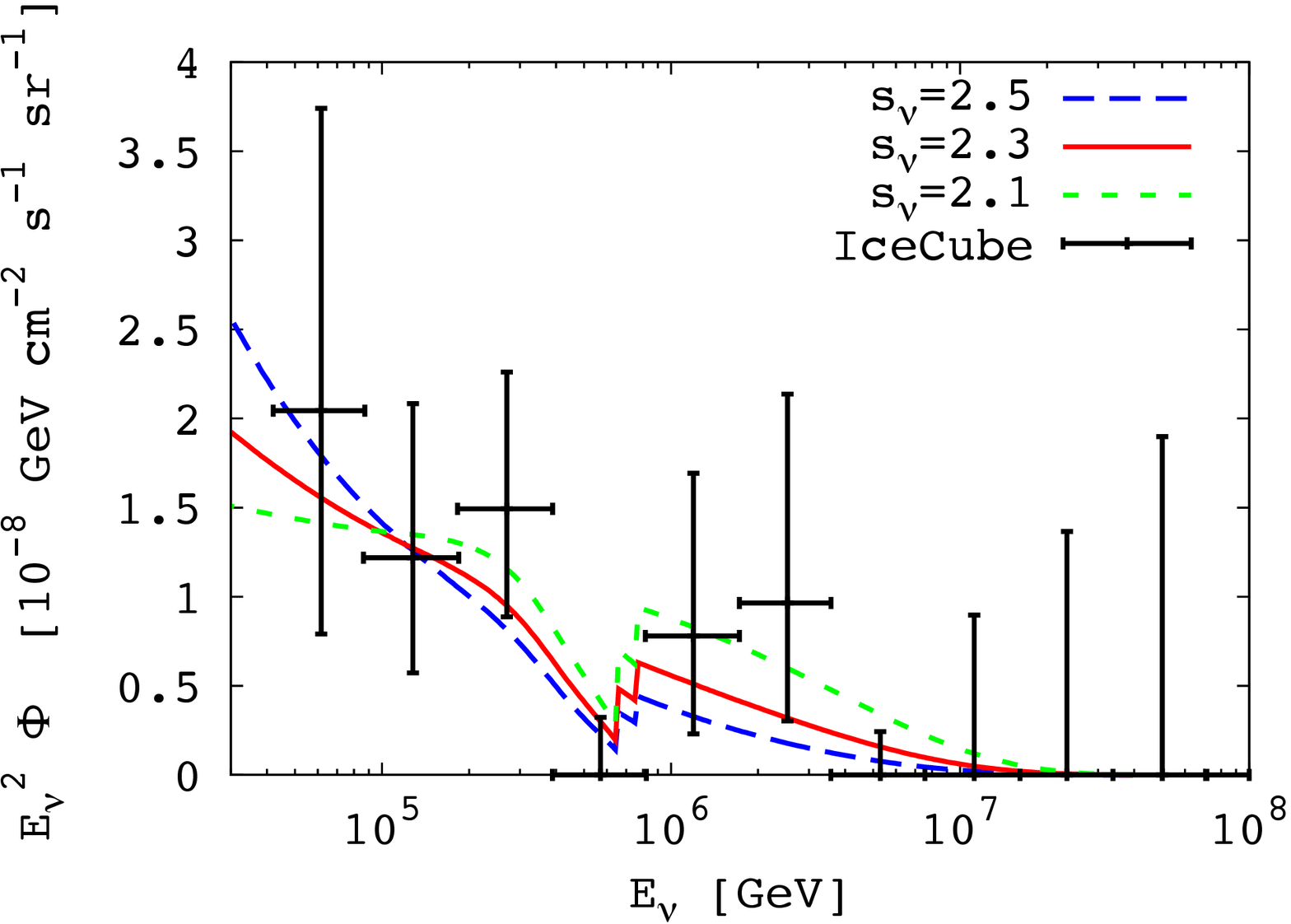}}
  \end{picture}
 \caption{Examples of the studies on secret neutrino interactions.
 [Left] The interaction between sterile neutrinos,
 which is mediated by a field with $M\lesssim 10$ MeV,
 suppresses the active-sterile mixing during the BBN
 era~\cite{Dasgupta:2013zpn}.
 [Right] The $L_{\mu} - L_{\tau}$ interaction
 mediated by the gauge field with a MeV scale mass can
 reproduce the gap in the cosmic neutrino spectrum
 and can simultaneously explain the discrepancy
 in the muon $g-2$~\cite{Araki:2015mya}.
 }
 \label{Fig:SecretNu}
\end{figure}

Finally I would like to discuss secret neutrino
interactions~\cite{BialynickaBirula:1964zz,Bardin:1970wq,Bilenky:1999dn}
--- neutrinos may communicate with the other invisible stuffs or
communicate with neutrinos through unknown forces.
In a broad sense, NSIs are also in this category, but here we are
more interested in the interactions mediated by
light fields (lighter than the electroweak scale).
They are recently discussed in various contexts.
For example, in Refs.~\cite{Dasgupta:2013zpn,Chu:2015ipa,Cherry:2016jol},
the authors introduced a new gauge force
that only sterile neutrino can feel.
Thanks to the force, the sterile neutrinos in the early universe
get the potential from thermal bath and the active-sterile mixing
is suppressed; See the left plot in Fig.~\ref{Fig:SecretNu}.
This helps the sterile neutrinos to escape the constraint from
the effective degree of freedom of the radiation in the era of
the Big Bang nucleosynthesis (BBN).

Here I would like to introduce a study on
the effect of a secret neutrino-neutrino
interaction on the cosmic neutrino spectrum.
In 2013, the IceCube collaboration announced
the first discovery of the high-energy cosmic neutrino events
with the energy
of peta electron volts~\cite{Aartsen:2013bka,Aartsen:2013jdh}.
After the discovery, they have accumulated the events,
and now we can know the spectrum of the cosmic neutrinos.
Although the statistics are still not enough
to conclude the shape of the spectrum,
it start showing some intriguing
features (Ref.~\cite{vanSantenICRC:2017} for 6-year data)
---
Non-power-law spectrum may be a hint to new physics. 
A popular idea is,
the peaks in the spectrum are made by
the decays of the dark matter fields~\cite{Feldstein:2013kka,Esmaili:2013gha,Bai:2013nga,Ema:2013nda,Ibarra:2013cra,Bhattacharya:2014vwa,Zavala:2014dla,Higaki:2014dwa,Bhattacharya:2014yha,Rott:2014kfa,Ema:2014ufa,Esmaili:2014rma,Roland:2015yoa,Berezhiani:2015fba,Murase:2015gea,Kopp:2015bfa,Esmaili:2015xpa,Ahlers:2015moa,Aisati:2015vma,Boucenna:2015tra,Anchordoqui:2015lqa,Ko:2015nma,Aisati:2015ova,Queiroz:2016zwd,Dev:2016qbd,DiBari:2016guw,Chianese:2016smc,Cohen:2016uyg,Garcia-Cely:2017oco,Borah:2017xgm,Hiroshima:2017hmy,Bhattacharya:2017jaw,ElAisati:2017ppn,Chianese:2017nwe}.
The other idea is,
the gap in the spectrum is made by a resonant scattering;
Originally, the spectrum follows a power-law, but the
cosmic neutrinos with a particular energy are resonantly
scattered with the cosmic neutrino background during their travel to the Earth
and lose their energies~\cite{Ioka:2014kca,Ng:2014pca,Ibe:2014pja,Blum:2014ewa,Araki:2014ona,Kamada:2015era,DiFranzo:2015qea,Araki:2015mya,Shoemaker:2015qul,Farzan:2016wym}.
In this scenario, the gap in the spectrum can be understood
as the signal of the new neutrino-neutrino interaction.
%
%
To scatter the sub-PeV cosmic neutrinos resonantly,
the new neutrino-neutrino interaction should be
mediated by a new field with a mass around the MeV scale.
The coupling of the interaction should be larger than $10^{-4}$
so that the cosmic neutrino can get the scattering
at least once during their travels of extra galactic distances$\sim$Gpc.
In the model with the $L_{\mu} - L_{\tau}$ gauge symmetry,
the relevant parameter region is still allowed, and
in addition, the parameter set can also explain the long-standing
discrepancy between theory and experiment in
the muon anomalous magnetic moment.
As shown in the right plot in Fig.~\ref{Fig:SecretNu},
the gap in the cosmic neutrino spectrum is reproduced 
with the model parameters that can explain the discrepancy
in the muon $g-2$.
Searches for the light $L_{\mu} - L_{\tau}$ gauge boson
have been discussed recently in 
Refs.~\cite{Essig:2013vha,Kaneta:2016uyt,Araki:2017wyg} (at lepton colliders),
\cite{Ibe:2016dir} (in meson decays),
\cite{Alekhin:2015byh,Gninenko:2014pea} (at beam dump experiments),
and \cite{Ge:2017poy} (in the trident process with atmospheric neutrinos).

\section{Summary}

In this talk, we have discussed the items listed below
\begin{itemize}
 \item Non-standard neutrino interactions
       
       NSIs as noise in parameter determinations.
       Three strategies to reduce the noise.

       NSIs as signal.
       Future sensitivities at long baseline experiments.

       Models for NSIs.
       There are still rooms for large NSIs with the help of
       light mediators. 
       
 \item Non-unitarity of the PMNS matrix

       Fits without the unitarity assumption.
       
       Current bounds in the MUV scheme.
       
       A way to test the unitarity triangle with neutrino oscillations.

 \item Secret neutrino interactions
       --- Interactions between neutrinos and invisibles
       mediated by light fields
       
       There are many studies in various contexts recently.
       We have a look at some examples.
       
\end{itemize}
We are not who we were 5 years ago ---
We know $\theta_{13}$ and are going to know the CP phase and
the mass hierarchy in the next years.
It is a good time to think over the plan for the future experiments,
taking the possibilities of exotics into consideration.
In this talk, I would like to emphasize that
it is important to measure oscillation patterns with high precision
and check if they are consistent with the standard oscillation scenario
without putting any assumption and bias.

%
\bibliographystyle{JHEP}
\bibliography{NuExotic}

\providecommand{\href}[2]{#2}\begingroup\raggedright\begin{thebibliography}{100}

\bibitem{Blennow:2016etl}
M.~Blennow, S.~Choubey, T.~Ohlsson, D.~Pramanik and S.~K. Raut, \emph{{A
  combined study of source, detector and matter non-standard neutrino
  interactions at DUNE}},
  \href{https://doi.org/10.1007/JHEP08(2016)090}{\emph{JHEP} {\bfseries 08}
  (2016) 090}, [\href{https://arxiv.org/abs/1606.08851}{{\ttfamily
  1606.08851}}].

\bibitem{Choubey:2015xha}
S.~Choubey, A.~Ghosh, T.~Ohlsson and D.~Tiwari, \emph{{Neutrino Physics with
  Non-Standard Interactions at INO}},
  \href{https://doi.org/10.1007/JHEP12(2015)126}{\emph{JHEP} {\bfseries 12}
  (2015) 126}, [\href{https://arxiv.org/abs/1507.02211}{{\ttfamily
  1507.02211}}].

\bibitem{Masud:2016gcl}
M.~Masud and P.~Mehta, \emph{{Nonstandard interactions and resolving the
  ordering of neutrino masses at DUNE and other long baseline experiments}},
  \href{https://doi.org/10.1103/PhysRevD.94.053007}{\emph{Phys. Rev.}
  {\bfseries D94} (2016) 053007},
  [\href{https://arxiv.org/abs/1606.05662}{{\ttfamily 1606.05662}}].

\bibitem{deGouvea:2016pom}
A.~de~Gouvea and K.~J. Kelly, \emph{{False Signals of CP-Invariance Violation
  at DUNE}},  \href{https://arxiv.org/abs/1605.09376}{{\ttfamily 1605.09376}}.

\bibitem{Agarwalla:2016fkh}
S.~K. Agarwalla, S.~S. Chatterjee and A.~Palazzo, \emph{{Degeneracy between
  $\theta_{23}$ octant and neutrino non-standard interactions at DUNE}},
  \href{https://doi.org/10.1016/j.physletb.2016.09.020}{\emph{Phys. Lett.}
  {\bfseries B762} (2016) 64--71},
  [\href{https://arxiv.org/abs/1607.01745}{{\ttfamily 1607.01745}}].

\bibitem{Dutta:2016czj}
D.~Dutta, P.~Ghoshal and S.~Roy, \emph{{Effect of Non Unitarity on Neutrino
  Mass Hierarchy determination at DUNE, NO$\nu$A and T2K}},
  \href{https://doi.org/10.1016/j.nuclphysb.2017.04.018}{\emph{Nucl. Phys.}
  {\bfseries B920} (2017) 385--401},
  [\href{https://arxiv.org/abs/1609.07094}{{\ttfamily 1609.07094}}].

\bibitem{Deepthi:2016erc}
K.~N. Deepthi, S.~Goswami and N.~Nath, \emph{{Can nonstandard interactions
  jeopardize the hierarchy sensitivity of DUNE?}},
  \href{https://doi.org/10.1103/PhysRevD.96.075023}{\emph{Phys. Rev.}
  {\bfseries D96} (2017) 075023},
  [\href{https://arxiv.org/abs/1612.00784}{{\ttfamily 1612.00784}}].

\bibitem{Masud:2015xva}
M.~Masud, A.~Chatterjee and P.~Mehta, \emph{{Probing CP violation signal at
  DUNE in presence of non-standard neutrino interactions}},
  \href{https://doi.org/10.1088/0954-3899/43/9/095005/meta,
  10.1088/0954-3899/43/9/095005}{\emph{J. Phys.} {\bfseries G43} (2016)
  095005}, [\href{https://arxiv.org/abs/1510.08261}{{\ttfamily 1510.08261}}].

\bibitem{Masud:2016bvp}
M.~Masud and P.~Mehta, \emph{{Nonstandard interactions spoiling the CP
  violation sensitivity at DUNE and other long baseline experiments}},
  \href{https://doi.org/10.1103/PhysRevD.94.013014}{\emph{Phys. Rev.}
  {\bfseries D94} (2016) 013014},
  [\href{https://arxiv.org/abs/1603.01380}{{\ttfamily 1603.01380}}].

\bibitem{Rout:2017udo}
J.~Rout, M.~Masud and P.~Mehta, \emph{{Can we probe intrinsic CP and T
  violations and nonunitarity at long baseline accelerator experiments?}},
  \href{https://doi.org/10.1103/PhysRevD.95.075035}{\emph{Phys. Rev.}
  {\bfseries D95} (2017) 075035},
  [\href{https://arxiv.org/abs/1702.02163}{{\ttfamily 1702.02163}}].

\bibitem{Miranda:2016wdr}
O.~G. Miranda, M.~Tortola and J.~W.~F. Valle, \emph{{New ambiguity in probing
  CP violation in neutrino oscillations}},
  \href{https://doi.org/10.1103/PhysRevLett.117.061804}{\emph{Phys. Rev. Lett.}
  {\bfseries 117} (2016) 061804},
  [\href{https://arxiv.org/abs/1604.05690}{{\ttfamily 1604.05690}}].

\bibitem{Ge:2016dlx}
S.-F. Ge and A.~{\relax Yu}. Smirnov, \emph{{Non-standard interactions and the
  CP phase measurements in neutrino oscillations at low energies}},
  \href{https://doi.org/10.1007/JHEP10(2016)138}{\emph{JHEP} {\bfseries 10}
  (2016) 138}, [\href{https://arxiv.org/abs/1607.08513}{{\ttfamily
  1607.08513}}].

\bibitem{Dutta:2016vcc}
D.~Dutta and P.~Ghoshal, \emph{{Probing CP violation with T2K, NO$\nu$A and
  DUNE in the presence of non-unitarity}},
  \href{https://doi.org/10.1007/JHEP09(2016)110}{\emph{JHEP} {\bfseries 09}
  (2016) 110}, [\href{https://arxiv.org/abs/1607.02500}{{\ttfamily
  1607.02500}}].

\bibitem{Dutta:2016eks}
D.~Dutta, P.~Ghoshal and S.~K. Sehrawat, \emph{{Octant of $\theta_{23}$ at long
  baseline neutrino experiments in the light of Non Unitary Leptonic mixing}},
  \href{https://doi.org/10.1103/PhysRevD.95.095007}{\emph{Phys. Rev.}
  {\bfseries D95} (2017) 095007},
  [\href{https://arxiv.org/abs/1610.07203}{{\ttfamily 1610.07203}}].

\bibitem{Das:2017fcz}
C.~R. Das, J.~Maalampi, J.~Pulido and S.~Vihonen, \emph{{Determination of the
  $\theta_{23}$ octant in long baseline neutrino experiments within and beyond
  the Standard Model}},  \href{https://arxiv.org/abs/1708.05182}{{\ttfamily
  1708.05182}}.

\bibitem{Bakhti:2016prn}
P.~Bakhti and Y.~Farzan, \emph{{CP-Violation and Non-Standard Interactions at
  the MOMENT}}, \href{https://doi.org/10.1007/JHEP07(2016)109}{\emph{JHEP}
  {\bfseries 07} (2016) 109},
  [\href{https://arxiv.org/abs/1602.07099}{{\ttfamily 1602.07099}}].

\bibitem{Cao:2014bea}
J.~Cao et~al., \emph{{Muon-decay medium-baseline neutrino beam facility}},
  \href{https://doi.org/10.1103/PhysRevSTAB.17.090101}{\emph{Phys. Rev. ST
  Accel. Beams} {\bfseries 17} (2014) 090101},
  [\href{https://arxiv.org/abs/1401.8125}{{\ttfamily 1401.8125}}].

\bibitem{Aberle:2013ssa}
C.~Aberle et~al., \emph{{Whitepaper on the DAEdALUS Program}},  in
  \emph{{Proceedings, 2013 Community Summer Study on the Future of U.S.
  Particle Physics: Snowmass on the Mississippi (CSS2013): Minneapolis, MN,
  USA, July 29-August 6, 2013}}, 2013,
  \href{https://arxiv.org/abs/1307.2949}{{\ttfamily 1307.2949}},
  \href{https://inspirehep.net/record/1242298/files/arXiv:1307.2949.pdf}{https://inspirehep.net/record/1242298/files/arXiv:1307.2949.pdf}.

\bibitem{Kyberd:2012iz}
{\scshape nuSTORM} collaboration, P.~Kyberd et~al., \emph{{nuSTORM - Neutrinos
  from STORed Muons: Letter of Intent to the Fermilab Physics Advisory
  Committee}},  \href{https://arxiv.org/abs/1206.0294}{{\ttfamily 1206.0294}}.

\bibitem{Adey:2013pio}
{\scshape nuSTORM} collaboration, D.~Adey et~al., \emph{{nuSTORM - Neutrinos
  from STORed Muons: Proposal to the Fermilab PAC}},
  \href{https://arxiv.org/abs/1308.6822}{{\ttfamily 1308.6822}}.

\bibitem{Baussan:2012cw}
E.~Baussan, M.~Dracos, T.~Ekelof, E.~F. Martinez, H.~Ohman and
  N.~Vassilopoulos, \emph{{The use the a high intensity neutrino beam from the
  ESS proton linac for measurement of neutrino CP violation and mass
  hierarchy}},  \href{https://arxiv.org/abs/1212.5048}{{\ttfamily 1212.5048}}.

\bibitem{Akimov:2017ade}
{\scshape COHERENT} collaboration, D.~Akimov et~al., \emph{{Observation of
  Coherent Elastic Neutrino-Nucleus Scattering}},
  \href{https://doi.org/10.1126/science.aao0990}{\emph{Science} {\bfseries 357}
  (2017) 1123--1126}, [\href{https://arxiv.org/abs/1708.01294}{{\ttfamily
  1708.01294}}].

\bibitem{Miranda:2004nb}
O.~G. Miranda, M.~A. Tortola and J.~W.~F. Valle, \emph{{Are solar neutrino
  oscillations robust?}},
  \href{https://doi.org/10.1088/1126-6708/2006/10/008}{\emph{JHEP} {\bfseries
  10} (2006) 008}, [\href{https://arxiv.org/abs/hep-ph/0406280}{{\ttfamily
  hep-ph/0406280}}].

\bibitem{Coloma:2017ncl}
P.~Coloma, M.~C. Gonzalez-Garcia, M.~Maltoni and T.~Schwetz, \emph{{A COHERENT
  enlightenment of the neutrino Dark Side}},
  \href{https://arxiv.org/abs/1708.02899}{{\ttfamily 1708.02899}}.

\bibitem{Liao:2017uzy}
J.~Liao and D.~Marfatia, \emph{{COHERENT constraints on nonstandard neutrino
  interactions}},
  \href{https://doi.org/10.1016/j.physletb.2017.10.046}{\emph{Phys. Lett.}
  {\bfseries B775} (2017) 54--57},
  [\href{https://arxiv.org/abs/1708.04255}{{\ttfamily 1708.04255}}].

\bibitem{Salvado:2016uqu}
J.~Salvado, O.~Mena, S.~Palomares-Ruiz and N.~Rius, \emph{{Non-standard
  interactions with high-energy atmospheric neutrinos at IceCube}},
  \href{https://doi.org/10.1007/JHEP01(2017)141}{\emph{JHEP} {\bfseries 01}
  (2017) 141}, [\href{https://arxiv.org/abs/1609.03450}{{\ttfamily
  1609.03450}}].

\bibitem{Aartsen:2017xtt}
{\scshape IceCube} collaboration, M.~G. Aartsen et~al., \emph{{Search for
  Nonstandard Neutrino Interactions with IceCube DeepCore}},
  \href{https://arxiv.org/abs/1709.07079}{{\ttfamily 1709.07079}}.

\bibitem{Coloma:2015kiu}
P.~Coloma, \emph{{Non-Standard Interactions in propagation at the Deep
  Underground Neutrino Experiment}},
  \href{https://doi.org/10.1007/JHEP03(2016)016}{\emph{JHEP} {\bfseries 03}
  (2016) 016}, [\href{https://arxiv.org/abs/1511.06357}{{\ttfamily
  1511.06357}}].

\bibitem{Fukasawa:2016nwn}
S.~Fukasawa and O.~Yasuda, \emph{{The possibility to observe the non-standard
  interaction by the Hyperkamiokande atmospheric neutrino experiment}},
  \href{https://doi.org/10.1016/j.nuclphysb.2016.11.004}{\emph{Nucl. Phys.}
  {\bfseries B914} (2017) 99--116},
  [\href{https://arxiv.org/abs/1608.05897}{{\ttfamily 1608.05897}}].

\bibitem{Kelly:2017kch}
K.~J. Kelly, \emph{{Searches for new physics at the Hyper-Kamiokande
  experiment}}, \href{https://doi.org/10.1103/PhysRevD.95.115009}{\emph{Phys.
  Rev.} {\bfseries D95} (2017) 115009},
  [\href{https://arxiv.org/abs/1703.00448}{{\ttfamily 1703.00448}}].

\bibitem{Coelho:2017cwp}
{\scshape KM3NeT} collaboration, J.~A.~B. Coelho, \emph{{Probing new physics
  with atmospheric neutrinos at KM3NeT-ORCA}},  in \emph{{27th International
  Conference on Neutrino Physics and Astrophysics (Neutrino 2016) London,
  United Kingdom, July 4-9, 2016}}, 2017,
  \href{https://arxiv.org/abs/1702.04508}{{\ttfamily 1702.04508}},
  \href{https://inspirehep.net/record/1513471/files/arXiv:1702.04508.pdf}{https://inspirehep.net/record/1513471/files/arXiv:1702.04508.pdf}.

\bibitem{Tang:2017qen}
J.~Tang and Y.~Zhang, \emph{{Study of Non-Standard Charged-Current Interactions
  at the MOMENT experiment}},
  \href{https://arxiv.org/abs/1705.09500}{{\ttfamily 1705.09500}}.

\bibitem{Blennow:2015nxa}
M.~Blennow, S.~Choubey, T.~Ohlsson and S.~K. Raut, \emph{{Exploring Source and
  Detector Non-Standard Neutrino Interactions at ESS$\nu$SB}},
  \href{https://doi.org/10.1007/JHEP09(2015)096}{\emph{JHEP} {\bfseries 09}
  (2015) 096}, [\href{https://arxiv.org/abs/1507.02868}{{\ttfamily
  1507.02868}}].

\bibitem{Liao:2017awz}
J.~Liao, D.~Marfatia and K.~Whisnant, \emph{{Nonstandard interactions in solar
  neutrino oscillations with Hyper-Kamiokande and JUNO}},
  \href{https://doi.org/10.1016/j.physletb.2017.05.054}{\emph{Phys. Lett.}
  {\bfseries B771} (2017) 247--253},
  [\href{https://arxiv.org/abs/1704.04711}{{\ttfamily 1704.04711}}].

\bibitem{Liao:2016bgf}
J.~Liao, D.~Marfatia and K.~Whisnant, \emph{{Nonmaximal neutrino mixing at
  NO$\nu$A from nonstandard interactions}},
  \href{https://doi.org/10.1016/j.physletb.2017.02.024}{\emph{Phys. Lett.}
  {\bfseries B767} (2017) 350--353},
  [\href{https://arxiv.org/abs/1609.01786}{{\ttfamily 1609.01786}}].

\bibitem{Fukasawa:2016gvm}
S.~Fukasawa, M.~Ghosh and O.~Yasuda, \emph{{Is nonstandard interaction a
  solution to the three neutrino tensions?}},
  \href{https://arxiv.org/abs/1609.04204}{{\ttfamily 1609.04204}}.

\bibitem{Grossman:1995wx}
Y.~Grossman, \emph{{Nonstandard neutrino interactions and neutrino oscillation
  experiments}},
  \href{https://doi.org/10.1016/0370-2693(95)01069-3}{\emph{Phys. Lett.}
  {\bfseries B359} (1995) 141--147},
  [\href{https://arxiv.org/abs/hep-ph/9507344}{{\ttfamily hep-ph/9507344}}].

\bibitem{Bergmann:1998ft}
S.~Bergmann and Y.~Grossman, \emph{{Can lepton flavor violating interactions
  explain the LSND results?}},
  \href{https://doi.org/10.1103/PhysRevD.59.093005}{\emph{Phys. Rev.}
  {\bfseries D59} (1999) 093005},
  [\href{https://arxiv.org/abs/hep-ph/9809524}{{\ttfamily hep-ph/9809524}}].

\bibitem{Bergmann:1999rz}
S.~Bergmann, Y.~Grossman and E.~Nardi, \emph{{Neutrino propagation in matter
  with general interactions}},
  \href{https://doi.org/10.1103/PhysRevD.60.093008}{\emph{Phys. Rev.}
  {\bfseries D60} (1999) 093008},
  [\href{https://arxiv.org/abs/hep-ph/9903517}{{\ttfamily hep-ph/9903517}}].

\bibitem{Bergmann:1999pk}
S.~Bergmann, Y.~Grossman and D.~M. Pierce, \emph{{Can lepton flavor violating
  interactions explain the atmospheric neutrino problem?}},
  \href{https://doi.org/10.1103/PhysRevD.61.053005}{\emph{Phys. Rev.}
  {\bfseries D61} (2000) 053005},
  [\href{https://arxiv.org/abs/hep-ph/9909390}{{\ttfamily hep-ph/9909390}}].

\bibitem{Berezhiani:2001rs}
Z.~Berezhiani and A.~Rossi, \emph{{Limits on the nonstandard interactions of
  neutrinos from e+ e- colliders}},
  \href{https://doi.org/10.1016/S0370-2693(02)01767-7}{\emph{Phys. Lett.}
  {\bfseries B535} (2002) 207--218},
  [\href{https://arxiv.org/abs/hep-ph/0111137}{{\ttfamily hep-ph/0111137}}].

\bibitem{Davidson:2003ha}
S.~Davidson, C.~Pena-Garay, N.~Rius and A.~Santamaria, \emph{{Present and
  future bounds on nonstandard neutrino interactions}},
  \href{https://doi.org/10.1088/1126-6708/2003/03/011}{\emph{JHEP} {\bfseries
  03} (2003) 011}, [\href{https://arxiv.org/abs/hep-ph/0302093}{{\ttfamily
  hep-ph/0302093}}].

\bibitem{Antusch:2008tz}
S.~Antusch, J.~P. Baumann and E.~Fernandez-Martinez, \emph{{Non-Standard
  Neutrino Interactions with Matter from Physics Beyond the Standard Model}},
  \href{https://doi.org/10.1016/j.nuclphysb.2008.11.018}{\emph{Nucl. Phys.}
  {\bfseries B810} (2009) 369--388},
  [\href{https://arxiv.org/abs/0807.1003}{{\ttfamily 0807.1003}}].

\bibitem{Gavela:2008ra}
M.~B. Gavela, D.~Hernandez, T.~Ota and W.~Winter, \emph{{Large gauge invariant
  non-standard neutrino interactions}},
  \href{https://doi.org/10.1103/PhysRevD.79.013007}{\emph{Phys. Rev.}
  {\bfseries D79} (2009) 013007},
  [\href{https://arxiv.org/abs/0809.3451}{{\ttfamily 0809.3451}}].

\bibitem{Biggio:2009kv}
C.~Biggio, M.~Blennow and E.~Fernandez-Martinez, \emph{{Loop bounds on
  non-standard neutrino interactions}},
  \href{https://doi.org/10.1088/1126-6708/2009/03/139}{\emph{JHEP} {\bfseries
  03} (2009) 139}, [\href{https://arxiv.org/abs/0902.0607}{{\ttfamily
  0902.0607}}].

\bibitem{Biggio:2009nt}
C.~Biggio, M.~Blennow and E.~Fernandez-Martinez, \emph{{General bounds on
  non-standard neutrino interactions}},
  \href{https://doi.org/10.1088/1126-6708/2009/08/090}{\emph{JHEP} {\bfseries
  08} (2009) 090}, [\href{https://arxiv.org/abs/0907.0097}{{\ttfamily
  0907.0097}}].

\bibitem{Farzan:2015doa}
Y.~Farzan, \emph{{A model for large non-standard interactions of neutrinos
  leading to the LMA-Dark solution}},
  \href{https://doi.org/10.1016/j.physletb.2015.07.015}{\emph{Phys. Lett.}
  {\bfseries B748} (2015) 311--315},
  [\href{https://arxiv.org/abs/1505.06906}{{\ttfamily 1505.06906}}].

\bibitem{Farzan:2015hkd}
Y.~Farzan and I.~M. Shoemaker, \emph{{Lepton Flavor Violating Non-Standard
  Interactions via Light Mediators}},
  \href{https://doi.org/10.1007/JHEP07(2016)033}{\emph{JHEP} {\bfseries 07}
  (2016) 033}, [\href{https://arxiv.org/abs/1512.09147}{{\ttfamily
  1512.09147}}].

\bibitem{Farzan:2016wym}
Y.~Farzan and J.~Heeck, \emph{{Neutrinophilic nonstandard interactions}},
  \href{https://doi.org/10.1103/PhysRevD.94.053010}{\emph{Phys. Rev.}
  {\bfseries D94} (2016) 053010},
  [\href{https://arxiv.org/abs/1607.07616}{{\ttfamily 1607.07616}}].

\bibitem{Machado:2016fwi}
P.~A.~N. Machado, \emph{{Flavor effects at the MeV and TeV scales}},
  \href{https://doi.org/10.1063/1.4953286}{\emph{AIP Conf. Proc.} {\bfseries
  1743} (2016) 030005}.

\bibitem{Babu:2017olk}
K.~S. Babu, A.~Friedland, P.~A.~N. Machado and I.~Mocioiu, \emph{{Flavor Gauge
  Models Below the Fermi Scale}},
  \href{https://arxiv.org/abs/1705.01822}{{\ttfamily 1705.01822}}.

\bibitem{Farzan:2015pca}
Y.~Farzan and S.~Hannestad, \emph{{Neutrinos secretly converting to lighter
  particles to please both KATRIN and the cosmos}},
  \href{https://doi.org/10.1088/1475-7516/2016/02/058}{\emph{JCAP} {\bfseries
  1602} (2016) 058}, [\href{https://arxiv.org/abs/1510.02201}{{\ttfamily
  1510.02201}}].

\bibitem{Huang:2017egl}
G.-y. Huang, T.~Ohlsson and S.~Zhou, \emph{{Observational Constraints on Secret
  Neutrino Interactions from Big Bang Nucleosynthesis}},
  \href{https://arxiv.org/abs/1712.04792}{{\ttfamily 1712.04792}}.

\bibitem{Datta:2017pfz}
A.~Datta, J.~Liao and D.~Marfatia, \emph{{A light $Z^\prime$ for the $R_K$
  puzzle and nonstandard neutrino interactions}},
  \href{https://doi.org/10.1016/j.physletb.2017.02.058}{\emph{Phys. Lett.}
  {\bfseries B768} (2017) 265--269},
  [\href{https://arxiv.org/abs/1702.01099}{{\ttfamily 1702.01099}}].

\bibitem{Parke:2015goa}
S.~Parke and M.~Ross-Lonergan, \emph{{Unitarity and the three flavor neutrino
  mixing matrix}},
  \href{https://doi.org/10.1103/PhysRevD.93.113009}{\emph{Phys. Rev.}
  {\bfseries D93} (2016) 113009},
  [\href{https://arxiv.org/abs/1508.05095}{{\ttfamily 1508.05095}}].

\bibitem{Fong:2016yyh}
C.~S. Fong, H.~Minakata and H.~Nunokawa, \emph{{A framework for testing
  leptonic unitarity by neutrino oscillation experiments}},
  \href{https://doi.org/10.1007/JHEP02(2017)114}{\emph{JHEP} {\bfseries 02}
  (2017) 114}, [\href{https://arxiv.org/abs/1609.08623}{{\ttfamily
  1609.08623}}].

\bibitem{Langacker:1988ur}
P.~Langacker and D.~London, \emph{{Mixing Between Ordinary and Exotic
  Fermions}}, \href{https://doi.org/10.1103/PhysRevD.38.886}{\emph{Phys. Rev.}
  {\bfseries D38} (1988) 886}.

\bibitem{Antusch:2006vwa}
S.~Antusch, C.~Biggio, E.~Fernandez-Martinez, M.~B. Gavela and J.~Lopez-Pavon,
  \emph{{Unitarity of the Leptonic Mixing Matrix}},
  \href{https://doi.org/10.1088/1126-6708/2006/10/084}{\emph{JHEP} {\bfseries
  10} (2006) 084}, [\href{https://arxiv.org/abs/hep-ph/0607020}{{\ttfamily
  hep-ph/0607020}}].

\bibitem{Berryman:2014yoa}
J.~M. Berryman, A.~de~Gouvea, D.~Hernandez and R.~L.~N. Oliveira,
  \emph{{Non-Unitary Neutrino Propagation From Neutrino Decay}},
  \href{https://doi.org/10.1016/j.physletb.2015.01.002}{\emph{Phys. Lett.}
  {\bfseries B742} (2015) 74--79},
  [\href{https://arxiv.org/abs/1407.6631}{{\ttfamily 1407.6631}}].

\bibitem{Coloma:2017zpg}
P.~Coloma and O.~L.~G. Peres, \emph{{Visible neutrino decay at DUNE}},
  \href{https://arxiv.org/abs/1705.03599}{{\ttfamily 1705.03599}}.

\bibitem{Choubey:2017dyu}
S.~Choubey, S.~Goswami and D.~Pramanik, \emph{{A Study of Invisible Neutrino
  Decay at DUNE and its Effects on $\theta_{23}$ Measurement}},
  \href{https://arxiv.org/abs/1705.05820}{{\ttfamily 1705.05820}}.

\bibitem{Escrihuela:2015wra}
F.~J. Escrihuela, D.~V. Forero, O.~G. Miranda, M.~Tortola and J.~W.~F. Valle,
  \emph{{On the description of nonunitary neutrino mixing}},
  \href{https://doi.org/10.1103/PhysRevD.93.119905,
  10.1103/PhysRevD.92.053009}{\emph{Phys. Rev.} {\bfseries D92} (2015) 053009},
  [\href{https://arxiv.org/abs/1503.08879}{{\ttfamily 1503.08879}}].

\bibitem{Ge:2016xya}
S.-F. Ge, P.~Pasquini, M.~Tortola and J.~W.~F. Valle, \emph{{Measuring the
  leptonic CP phase in neutrino oscillations with nonunitary mixing}},
  \href{https://doi.org/10.1103/PhysRevD.95.033005}{\emph{Phys. Rev.}
  {\bfseries D95} (2017) 033005},
  [\href{https://arxiv.org/abs/1605.01670}{{\ttfamily 1605.01670}}].

\bibitem{Escrihuela:2016ube}
F.~J. Escrihuela, D.~V. Forero, O.~G. Miranda, M.~Tortola and J.~W.~F. Valle,
  \emph{{Probing CP violation with non-unitary mixing in long-baseline neutrino
  oscillation experiments: DUNE as a case study}},
  \href{https://doi.org/10.1088/1367-2630/aa79ec}{\emph{New J. Phys.}
  {\bfseries 19} (2017) 093005},
  [\href{https://arxiv.org/abs/1612.07377}{{\ttfamily 1612.07377}}].

\bibitem{Blennow:2016jkn}
M.~Blennow, P.~Coloma, E.~Fernandez-Martinez, J.~Hernandez-Garcia and
  J.~Lopez-Pavon, \emph{{Non-Unitarity, sterile neutrinos, and Non-Standard
  neutrino Interactions}},
  \href{https://doi.org/10.1007/JHEP04(2017)153}{\emph{JHEP} {\bfseries 04}
  (2017) 153}, [\href{https://arxiv.org/abs/1609.08637}{{\ttfamily
  1609.08637}}].

\bibitem{Antusch:2014woa}
S.~Antusch and O.~Fischer, \emph{{Non-unitarity of the leptonic mixing matrix:
  Present bounds and future sensitivities}},
  \href{https://doi.org/10.1007/JHEP10(2014)094}{\emph{JHEP} {\bfseries 10}
  (2014) 094}, [\href{https://arxiv.org/abs/1407.6607}{{\ttfamily 1407.6607}}].

\bibitem{Antusch:2016brq}
S.~Antusch and O.~Fischer, \emph{{Probing the nonunitarity of the leptonic
  mixing matrix at the CEPC}},
  \href{https://doi.org/10.1142/S0217751X16440061}{\emph{Int. J. Mod. Phys.}
  {\bfseries A31} (2016) 1644006},
  [\href{https://arxiv.org/abs/1604.00208}{{\ttfamily 1604.00208}}].

\bibitem{Fernandez-Martinez:2016lgt}
E.~Fernandez-Martinez, J.~Hernandez-Garcia and J.~Lopez-Pavon, \emph{{Global
  constraints on heavy neutrino mixing}},
  \href{https://doi.org/10.1007/JHEP08(2016)033}{\emph{JHEP} {\bfseries 08}
  (2016) 033}, [\href{https://arxiv.org/abs/1605.08774}{{\ttfamily
  1605.08774}}].

\bibitem{Meloni:2009cg}
D.~Meloni, T.~Ohlsson, W.~Winter and H.~Zhang, \emph{{Non-standard interactions
  versus non-unitary lepton flavor mixing at a neutrino factory}},
  \href{https://doi.org/10.1007/JHEP04(2010)041}{\emph{JHEP} {\bfseries 04}
  (2010) 041}, [\href{https://arxiv.org/abs/0912.2735}{{\ttfamily 0912.2735}}].

\bibitem{Tang:2017khg}
J.~Tang, Y.~Zhang and Y.-F. Li, \emph{{Probing Direct and Indirect Unitarity
  Violation in Future Accelerator Neutrino Facilities}},
  \href{https://doi.org/10.1016/j.physletb.2017.09.055}{\emph{Phys. Lett.}
  {\bfseries B774} (2017) 217--224},
  [\href{https://arxiv.org/abs/1708.04909}{{\ttfamily 1708.04909}}].

\bibitem{Sato:2000wv}
J.~Sato, \emph{{Neutrino oscillation and CP violation}},
  \href{https://doi.org/10.1016/S0168-9002(01)01287-6}{\emph{Nucl. Instrum.
  Meth.} {\bfseries A472} (2001) 434--439},
  [\href{https://arxiv.org/abs/hep-ph/0008056}{{\ttfamily hep-ph/0008056}}].

\bibitem{Farzan:2002ct}
Y.~Farzan and A.~{\relax Yu}. Smirnov, \emph{{Leptonic unitarity triangle and
  CP violation}}, \href{https://doi.org/10.1103/PhysRevD.65.113001}{\emph{Phys.
  Rev.} {\bfseries D65} (2002) 113001},
  [\href{https://arxiv.org/abs/hep-ph/0201105}{{\ttfamily hep-ph/0201105}}].

\bibitem{Pas:2016qbg}
H.~Paes and P.~Sicking, \emph{{Discriminating sterile neutrinos and unitarity
  violation with CP invariants}},
  \href{https://doi.org/10.1103/PhysRevD.95.075004}{\emph{Phys. Rev.}
  {\bfseries D95} (2017) 075004},
  [\href{https://arxiv.org/abs/1611.08450}{{\ttfamily 1611.08450}}].

\bibitem{Jarlskog:1985ht}
C.~Jarlskog, \emph{{Commutator of the Quark Mass Matrices in the Standard
  Electroweak Model and a Measure of Maximal CP Violation}},
  \href{https://doi.org/10.1103/PhysRevLett.55.1039}{\emph{Phys. Rev. Lett.}
  {\bfseries 55} (1985) 1039}.

\bibitem{Jarlskog:1985cw}
C.~Jarlskog, \emph{{A Basis Independent Formulation of the Connection Between
  Quark Mass Matrices, CP Violation and Experiment}},
  \href{https://doi.org/10.1007/BF01565198}{\emph{Z. Phys.} {\bfseries C29}
  (1985) 491--497}.

\bibitem{Dasgupta:2013zpn}
B.~Dasgupta and J.~Kopp, \emph{{Cosmologically Safe eV-Scale Sterile Neutrinos
  and Improved Dark Matter Structure}},
  \href{https://doi.org/10.1103/PhysRevLett.112.031803}{\emph{Phys. Rev. Lett.}
  {\bfseries 112} (2014) 031803},
  [\href{https://arxiv.org/abs/1310.6337}{{\ttfamily 1310.6337}}].

\bibitem{Araki:2015mya}
T.~Araki, F.~Kaneko, T.~Ota, J.~Sato and T.~Shimomura, \emph{{MeV scale
  leptonic force for cosmic neutrino spectrum and muon anomalous magnetic
  moment}}, \href{https://doi.org/10.1103/PhysRevD.93.013014}{\emph{Phys. Rev.}
  {\bfseries D93} (2016) 013014},
  [\href{https://arxiv.org/abs/1508.07471}{{\ttfamily 1508.07471}}].

\bibitem{BialynickaBirula:1964zz}
Z.~Bialynicka-Birula, \emph{{Do Neutrinos Interact between Themselves?}},
  \href{https://doi.org/10.1007/BF02749481}{\emph{Nuovo Cim.} {\bfseries 33}
  (1964) 1484--1487}.

\bibitem{Bardin:1970wq}
D.~{\relax Yu}. Bardin, S.~M. Bilenky and B.~Pontecorvo, \emph{{On the nu - nu
  interaction}},
  \href{https://doi.org/10.1016/0370-2693(70)90602-7}{\emph{Phys. Lett.}
  {\bfseries 32B} (1970) 121--124}.

\bibitem{Bilenky:1999dn}
M.~S. Bilenky and A.~Santamaria, \emph{{'Secret' neutrino interactions}},  in
  \emph{{Neutrino mixing. Festschrift in honour of Samoil Bilenky's 70th
  birthday. Proceedings, International Meeting, Turin, Italy, March 25-27,
  1999}}, pp.~50--61, 1999,
  \href{https://arxiv.org/abs/hep-ph/9908272}{{\ttfamily hep-ph/9908272}}.

\bibitem{Chu:2015ipa}
X.~Chu, B.~Dasgupta and J.~Kopp, \emph{{Sterile neutrinos with secret
  interactions—lasting friendship with cosmology}},
  \href{https://doi.org/10.1088/1475-7516/2015/10/011}{\emph{JCAP} {\bfseries
  1510} (2015) 011}, [\href{https://arxiv.org/abs/1505.02795}{{\ttfamily
  1505.02795}}].

\bibitem{Cherry:2016jol}
J.~F. Cherry, A.~Friedland and I.~M. Shoemaker, \emph{{Short-baseline neutrino
  oscillations, Planck, and IceCube}},
  \href{https://arxiv.org/abs/1605.06506}{{\ttfamily 1605.06506}}.

\bibitem{Aartsen:2013bka}
{\scshape IceCube Collaboration} collaboration, M.~Aartsen et~al., \emph{{First
  observation of PeV-energy neutrinos with IceCube}},
  \href{https://doi.org/10.1103/PhysRevLett.111.021103}{\emph{Phys.Rev.Lett.}
  {\bfseries 111} (2013) 021103},
  [\href{https://arxiv.org/abs/1304.5356}{{\ttfamily 1304.5356}}].

\bibitem{Aartsen:2013jdh}
{\scshape IceCube} collaboration, M.~Aartsen et~al., \emph{{Evidence for
  High-Energy Extraterrestrial Neutrinos at the IceCube Detector}},
  \href{https://doi.org/10.1126/science.1242856}{\emph{Science} {\bfseries 342}
  (2013) 1242856}, [\href{https://arxiv.org/abs/1311.5238}{{\ttfamily
  1311.5238}}].

\bibitem{vanSantenICRC:2017}
{\scshape IceCube} collaboration, J.~van Santen, \emph{{Highlights from IceCube
  and prospects for IceCube-Gen2 at the 35th International Cosmic Ray
  Conference (ICRC2017)}}, .

\bibitem{Feldstein:2013kka}
B.~Feldstein, A.~Kusenko, S.~Matsumoto and T.~T. Yanagida, \emph{{Neutrinos at
  IceCube from Heavy Decaying Dark Matter}},
  \href{https://doi.org/10.1103/PhysRevD.88.015004}{\emph{Phys.Rev.} {\bfseries
  D88} (2013) 015004}, [\href{https://arxiv.org/abs/1303.7320}{{\ttfamily
  1303.7320}}].

\bibitem{Esmaili:2013gha}
A.~Esmaili and P.~D. Serpico, \emph{{Are IceCube neutrinos unveiling PeV-scale
  decaying dark matter?}},
  \href{https://doi.org/10.1088/1475-7516/2013/11/054}{\emph{JCAP} {\bfseries
  1311} (2013) 054}, [\href{https://arxiv.org/abs/1308.1105}{{\ttfamily
  1308.1105}}].

\bibitem{Bai:2013nga}
Y.~Bai, R.~Lu and J.~Salvado, \emph{{Geometric Compatibility of IceCube TeV-PeV
  Neutrino Excess and its Galactic Dark Matter Origin}},
  \href{https://doi.org/10.1007/JHEP01(2016)161}{\emph{JHEP} {\bfseries 01}
  (2016) 161}, [\href{https://arxiv.org/abs/1311.5864}{{\ttfamily 1311.5864}}].

\bibitem{Ema:2013nda}
Y.~Ema, R.~Jinno and T.~Moroi, \emph{{Cosmic-Ray Neutrinos from the Decay of
  Long-Lived Particle and the Recent IceCube Result}},
  \href{https://doi.org/10.1016/j.physletb.2014.04.021}{\emph{Phys. Lett.}
  {\bfseries B733} (2014) 120--125},
  [\href{https://arxiv.org/abs/1312.3501}{{\ttfamily 1312.3501}}].

\bibitem{Ibarra:2013cra}
A.~Ibarra, D.~Tran and C.~Weniger, \emph{{Indirect Searches for Decaying Dark
  Matter}},
  \href{https://doi.org/10.1142/S0217751X13300408}{\emph{Int.J.Mod.Phys.}
  {\bfseries A28} (2013) 1330040},
  [\href{https://arxiv.org/abs/1307.6434}{{\ttfamily 1307.6434}}].

\bibitem{Bhattacharya:2014vwa}
A.~Bhattacharya, M.~H. Reno and I.~Sarcevic, \emph{{Reconciling neutrino flux
  from heavy dark matter decay and recent events at IceCube}},
  \href{https://doi.org/10.1007/JHEP06(2014)110}{\emph{JHEP} {\bfseries 06}
  (2014) 110}, [\href{https://arxiv.org/abs/1403.1862}{{\ttfamily 1403.1862}}].

\bibitem{Zavala:2014dla}
J.~Zavala, \emph{{Galactic PeV neutrinos from dark matter annihilation}},
  \href{https://doi.org/10.1103/PhysRevD.89.123516}{\emph{Phys. Rev.}
  {\bfseries D89} (2014) 123516},
  [\href{https://arxiv.org/abs/1404.2932}{{\ttfamily 1404.2932}}].

\bibitem{Higaki:2014dwa}
T.~Higaki, R.~Kitano and R.~Sato, \emph{{Neutrinoful Universe}},
  \href{https://doi.org/10.1007/JHEP07(2014)044}{\emph{JHEP} {\bfseries 07}
  (2014) 044}, [\href{https://arxiv.org/abs/1405.0013}{{\ttfamily 1405.0013}}].

\bibitem{Bhattacharya:2014yha}
A.~Bhattacharya, R.~Gandhi and A.~Gupta, \emph{{The Direct Detection of Boosted
  Dark Matter at High Energies and PeV events at IceCube}},
  \href{https://doi.org/10.1088/1475-7516/2015/03/027}{\emph{JCAP} {\bfseries
  1503} (2015) 027}, [\href{https://arxiv.org/abs/1407.3280}{{\ttfamily
  1407.3280}}].

\bibitem{Rott:2014kfa}
C.~Rott, K.~Kohri and S.~C. Park, \emph{{Superheavy dark matter and IceCube
  neutrino signals: Bounds on decaying dark matter}},
  \href{https://doi.org/10.1103/PhysRevD.92.023529}{\emph{Phys. Rev.}
  {\bfseries D92} (2015) 023529},
  [\href{https://arxiv.org/abs/1408.4575}{{\ttfamily 1408.4575}}].

\bibitem{Ema:2014ufa}
Y.~Ema, R.~Jinno and T.~Moroi, \emph{{Cosmological Implications of High-Energy
  Neutrino Emission from the Decay of Long-Lived Particle}},
  \href{https://doi.org/10.1007/JHEP10(2014)150}{\emph{JHEP} {\bfseries 10}
  (2014) 150}, [\href{https://arxiv.org/abs/1408.1745}{{\ttfamily 1408.1745}}].

\bibitem{Esmaili:2014rma}
A.~Esmaili, S.~K. Kang and P.~D. Serpico, \emph{{IceCube events and decaying
  dark matter: hints and constraints}},
  \href{https://doi.org/10.1088/1475-7516/2014/12/054}{\emph{JCAP} {\bfseries
  1412} (2014) 054}, [\href{https://arxiv.org/abs/1410.5979}{{\ttfamily
  1410.5979}}].

\bibitem{Roland:2015yoa}
S.~B. Roland, B.~Shakya and J.~D. Wells, \emph{{PeV neutrinos and a 3.5 keV
  x-ray line from a PeV-scale supersymmetric neutrino sector}},
  \href{https://doi.org/10.1103/PhysRevD.92.095018}{\emph{Phys. Rev.}
  {\bfseries D92} (2015) 095018},
  [\href{https://arxiv.org/abs/1506.08195}{{\ttfamily 1506.08195}}].

\bibitem{Berezhiani:2015fba}
Z.~Berezhiani, \emph{{Shadow dark matter, sterile neutrinos and neutrino events
  at IceCube, Neutrino Oscillation Workshop (NOW 2014) Conca Specchiulla,
  Otranto, Lecce, Italy, September 7-14, 2014}},
  \href{https://arxiv.org/abs/1506.09040}{{\ttfamily 1506.09040}}.

\bibitem{Murase:2015gea}
K.~Murase, R.~Laha, S.~Ando and M.~Ahlers, \emph{{Testing the Dark Matter
  Scenario for PeV Neutrinos Observed in IceCube}},
  \href{https://doi.org/10.1103/PhysRevLett.115.071301}{\emph{Phys. Rev. Lett.}
  {\bfseries 115} (2015) 071301},
  [\href{https://arxiv.org/abs/1503.04663}{{\ttfamily 1503.04663}}].

\bibitem{Kopp:2015bfa}
J.~Kopp, J.~Liu and X.-P. Wang, \emph{{Boosted Dark Matter in IceCube and at
  the Galactic Center}},
  \href{https://doi.org/10.1007/JHEP04(2015)105}{\emph{JHEP} {\bfseries 04}
  (2015) 105}, [\href{https://arxiv.org/abs/1503.02669}{{\ttfamily
  1503.02669}}].

\bibitem{Esmaili:2015xpa}
A.~Esmaili and P.~D. Serpico, \emph{{Gamma-ray bounds from EAS detectors and
  heavy decaying dark matter constraints}},
  \href{https://doi.org/10.1088/1475-7516/2015/10/014}{\emph{JCAP} {\bfseries
  1510} (2015) 014}, [\href{https://arxiv.org/abs/1505.06486}{{\ttfamily
  1505.06486}}].

\bibitem{Ahlers:2015moa}
M.~Ahlers, Y.~Bai, V.~Barger and R.~Lu, \emph{{Galactic neutrinos in the TeV to
  PeV range}}, \href{https://doi.org/10.1103/PhysRevD.93.013009}{\emph{Phys.
  Rev.} {\bfseries D93} (2016) 013009},
  [\href{https://arxiv.org/abs/1505.03156}{{\ttfamily 1505.03156}}].

\bibitem{Aisati:2015vma}
C.~El~Aisati, M.~Gustafsson and T.~Hambye, \emph{{New Search for Monochromatic
  Neutrinos from Dark Matter Decay}},
  \href{https://doi.org/10.1103/PhysRevD.92.123515}{\emph{Phys. Rev.}
  {\bfseries D92} (2015) 123515},
  [\href{https://arxiv.org/abs/1506.02657}{{\ttfamily 1506.02657}}].

\bibitem{Boucenna:2015tra}
S.~M. Boucenna, M.~Chianese, G.~Mangano, G.~Miele, S.~Morisi, O.~Pisanti
  et~al., \emph{{Decaying Leptophilic Dark Matter at IceCube}},
  \href{https://doi.org/10.1088/1475-7516/2015/12/055}{\emph{JCAP} {\bfseries
  1512} (2015) 055}, [\href{https://arxiv.org/abs/1507.01000}{{\ttfamily
  1507.01000}}].

\bibitem{Anchordoqui:2015lqa}
L.~A. Anchordoqui, V.~Barger, H.~Goldberg, X.~Huang, D.~Marfatia, L.~H.~M.
  da~Silva et~al., \emph{{IceCube neutrinos, decaying dark matter, and the
  Hubble constant}}, \href{https://doi.org/10.1103/PhysRevD.92.061301,
  10.1103/PhysRevD.94.069901}{\emph{Phys. Rev.} {\bfseries D92} (2015) 061301},
  [\href{https://arxiv.org/abs/1506.08788}{{\ttfamily 1506.08788}}].

\bibitem{Ko:2015nma}
P.~Ko and Y.~Tang, \emph{{IceCube Events from Heavy DM decays through the
  Right-handed Neutrino Portal}},
  \href{https://doi.org/10.1016/j.physletb.2015.10.021}{\emph{Phys. Lett.}
  {\bfseries B751} (2015) 81--88},
  [\href{https://arxiv.org/abs/1508.02500}{{\ttfamily 1508.02500}}].

\bibitem{Aisati:2015ova}
C.~El~Aisati, M.~Gustafsson, T.~Hambye and T.~Scarna, \emph{{Dark Matter Decay
  to a Photon and a Neutrino: the Double Monochromatic Smoking Gun Scenario}},
  \href{https://doi.org/10.1103/PhysRevD.93.043535}{\emph{Phys. Rev.}
  {\bfseries D93} (2016) 043535},
  [\href{https://arxiv.org/abs/1510.05008}{{\ttfamily 1510.05008}}].

\bibitem{Queiroz:2016zwd}
F.~S. Queiroz, C.~E. Yaguna and C.~Weniger, \emph{{Gamma-ray Limits on Neutrino
  Lines}}, \href{https://doi.org/10.1088/1475-7516/2016/05/050}{\emph{JCAP}
  {\bfseries 1605} (2016) 050},
  [\href{https://arxiv.org/abs/1602.05966}{{\ttfamily 1602.05966}}].

\bibitem{Dev:2016qbd}
P.~S.~B. Dev, D.~Kazanas, R.~N. Mohapatra, V.~L. Teplitz and Y.~Zhang,
  \emph{{Heavy right-handed neutrino dark matter and PeV neutrinos at
  IceCube}}, \href{https://doi.org/10.1088/1475-7516/2016/08/034}{\emph{JCAP}
  {\bfseries 1608} (2016) 034},
  [\href{https://arxiv.org/abs/1606.04517}{{\ttfamily 1606.04517}}].

\bibitem{DiBari:2016guw}
P.~Di~Bari, P.~O. Ludl and S.~Palomares-Ruiz, \emph{{Unifying leptogenesis,
  dark matter and high-energy neutrinos with right-handed neutrino mixing via
  Higgs portal}},
  \href{https://doi.org/10.1088/1475-7516/2016/11/044}{\emph{JCAP} {\bfseries
  1611} (2016) 044}, [\href{https://arxiv.org/abs/1606.06238}{{\ttfamily
  1606.06238}}].

\bibitem{Chianese:2016smc}
M.~Chianese and A.~Merle, \emph{{A Consistent Theory of Decaying Dark Matter
  Connecting IceCube to the Sesame Street}},
  \href{https://doi.org/10.1088/1475-7516/2017/04/017}{\emph{JCAP} {\bfseries
  1704} (2017) 017}, [\href{https://arxiv.org/abs/1607.05283}{{\ttfamily
  1607.05283}}].

\bibitem{Cohen:2016uyg}
T.~Cohen, K.~Murase, N.~L. Rodd, B.~R. Safdi and Y.~Soreq, \emph{{$\gamma$-ray
  Constraints on Decaying Dark Matter and Implications for IceCube}},
  \href{https://doi.org/10.1103/PhysRevLett.119.021102}{\emph{Phys. Rev. Lett.}
  {\bfseries 119} (2017) 021102},
  [\href{https://arxiv.org/abs/1612.05638}{{\ttfamily 1612.05638}}].

\bibitem{Garcia-Cely:2017oco}
C.~Garcia-Cely and J.~Heeck, \emph{{Neutrino Lines from Majoron Dark Matter}},
  \href{https://doi.org/10.1007/JHEP05(2017)102}{\emph{JHEP} {\bfseries 05}
  (2017) 102}, [\href{https://arxiv.org/abs/1701.07209}{{\ttfamily
  1701.07209}}].

\bibitem{Borah:2017xgm}
D.~Borah, A.~Dasgupta, U.~K. Dey, S.~Patra and G.~Tomar, \emph{{Multi-component
  Fermionic Dark Matter and IceCube PeV scale Neutrinos in Left-Right Model
  with Gauge Unification}},
  \href{https://doi.org/10.1007/JHEP09(2017)005}{\emph{JHEP} {\bfseries 09}
  (2017) 005}, [\href{https://arxiv.org/abs/1704.04138}{{\ttfamily
  1704.04138}}].

\bibitem{Hiroshima:2017hmy}
N.~Hiroshima, R.~Kitano, K.~Kohri and K.~Murase, \emph{{High-energy Neutrinos
  from Multi-body Decaying Dark Matter}},
  \href{https://arxiv.org/abs/1705.04419}{{\ttfamily 1705.04419}}.

\bibitem{Bhattacharya:2017jaw}
A.~Bhattacharya, A.~Esmaili, S.~Palomares-Ruiz and I.~Sarcevic, \emph{{Probing
  decaying heavy dark matter with the 4-year IceCube HESE data}},
  \href{https://doi.org/10.1088/1475-7516/2017/07/027}{\emph{JCAP} {\bfseries
  1707} (2017) 027}, [\href{https://arxiv.org/abs/1706.05746}{{\ttfamily
  1706.05746}}].

\bibitem{ElAisati:2017ppn}
C.~El~Aisati, C.~Garcia-Cely, T.~Hambye and L.~Vanderheyden, \emph{{Prospects
  for discovering a neutrino line induced by dark matter annihilation}},
  \href{https://doi.org/10.1088/1475-7516/2017/10/021}{\emph{JCAP} {\bfseries
  1710} (2017) 021}, [\href{https://arxiv.org/abs/1706.06600}{{\ttfamily
  1706.06600}}].

\bibitem{Chianese:2017nwe}
M.~Chianese, G.~Miele and S.~Morisi, \emph{{Interpreting IceCube 6-year HESE
  data as an evidence for hundred TeV decaying Dark Matter}},
  \href{https://doi.org/10.1016/j.physletb.2017.09.016}{\emph{Phys. Lett.}
  {\bfseries B773} (2017) 591--595},
  [\href{https://arxiv.org/abs/1707.05241}{{\ttfamily 1707.05241}}].

\bibitem{Ioka:2014kca}
K.~Ioka and K.~Murase, \emph{{IceCube PeV-EeV neutrinos and secret interactions
  of neutrinos}}, \href{https://doi.org/10.1093/ptep/ptu090}{\emph{PTEP}
  {\bfseries 2014} (2014) 061E01},
  [\href{https://arxiv.org/abs/1404.2279}{{\ttfamily 1404.2279}}].

\bibitem{Ng:2014pca}
K.~C.~Y. Ng and J.~F. Beacom, \emph{{Cosmic neutrino cascades from secret
  neutrino interactions}},
  \href{https://doi.org/10.1103/PhysRevD.90.065035}{\emph{Phys.Rev.} {\bfseries
  D90} (2014) 065035}, [\href{https://arxiv.org/abs/1404.2288}{{\ttfamily
  1404.2288}}].

\bibitem{Ibe:2014pja}
M.~Ibe and K.~Kaneta, \emph{{C$\nu$ B absorption line in the neutrino spectrum
  at IceCube}},
  \href{https://doi.org/10.1103/PhysRevD.90.053011}{\emph{Phys.Rev.} {\bfseries
  D90} (2014) 053011}, [\href{https://arxiv.org/abs/1407.2848}{{\ttfamily
  1407.2848}}].

\bibitem{Blum:2014ewa}
K.~Blum, A.~Hook and K.~Murase, \emph{{High energy neutrino telescopes as a
  probe of the neutrino mass mechanism}},
  \href{https://arxiv.org/abs/1408.3799}{{\ttfamily 1408.3799}}.

\bibitem{Araki:2014ona}
T.~Araki, F.~Kaneko, Y.~Konishi, T.~Ota, J.~Sato et~al., \emph{{Cosmic neutrino
  spectrum and the muon anomalous magnetic moment in the gauged
  $L_{\mu}-L_{\tau}$ model}},
  \href{https://doi.org/10.1103/PhysRevD.91.037301}{\emph{Phys.Rev.} {\bfseries
  D91} (2015) 037301}, [\href{https://arxiv.org/abs/1409.4180}{{\ttfamily
  1409.4180}}].

\bibitem{Kamada:2015era}
A.~Kamada and H.-B. Yu, \emph{{Coherent Propagation of PeV Neutrinos and the
  Dip in the Neutrino Spectrum at IceCube}},
  \href{https://doi.org/10.1103/PhysRevD.92.113004}{\emph{Phys. Rev.}
  {\bfseries D92} (2015) 113004},
  [\href{https://arxiv.org/abs/1504.00711}{{\ttfamily 1504.00711}}].

\bibitem{DiFranzo:2015qea}
A.~DiFranzo and D.~Hooper, \emph{{Searching for MeV-Scale Gauge Bosons with
  IceCube}}, \href{https://doi.org/10.1103/PhysRevD.92.095007}{\emph{Phys.
  Rev.} {\bfseries D92} (2015) 095007},
  [\href{https://arxiv.org/abs/1507.03015}{{\ttfamily 1507.03015}}].

\bibitem{Shoemaker:2015qul}
I.~M. Shoemaker and K.~Murase, \emph{{Probing BSM Neutrino Physics with Flavor
  and Spectral Distortions: Prospects for Future High-Energy Neutrino
  Telescopes}}, \href{https://doi.org/10.1103/PhysRevD.93.085004}{\emph{Phys.
  Rev.} {\bfseries D93} (2016) 085004},
  [\href{https://arxiv.org/abs/1512.07228}{{\ttfamily 1512.07228}}].

\bibitem{Essig:2013vha}
R.~Essig, J.~Mardon, M.~Papucci, T.~Volansky and Y.-M. Zhong,
  \emph{{Constraining Light Dark Matter with Low-Energy $e^+e^-$ Colliders}},
  \href{https://doi.org/10.1007/JHEP11(2013)167}{\emph{JHEP} {\bfseries 11}
  (2013) 167}, [\href{https://arxiv.org/abs/1309.5084}{{\ttfamily 1309.5084}}].

\bibitem{Kaneta:2016uyt}
Y.~Kaneta and T.~Shimomura, \emph{{On the possibility of a search for the
  $L_\mu - L_\tau$ gauge boson at Belle-II and neutrino beam experiments}},
  \href{https://doi.org/10.1093/ptep/ptx050}{\emph{PTEP} {\bfseries 2017}
  (2017) 053B04}, [\href{https://arxiv.org/abs/1701.00156}{{\ttfamily
  1701.00156}}].

\bibitem{Araki:2017wyg}
T.~Araki, S.~Hoshino, T.~Ota, J.~Sato and T.~Shimomura, \emph{{Detecting the
  $L_{\mu}-L_{\tau}$ gauge boson at Belle II}},
  \href{https://doi.org/10.1103/PhysRevD.95.055006}{\emph{Phys. Rev.}
  {\bfseries D95} (2017) 055006},
  [\href{https://arxiv.org/abs/1702.01497}{{\ttfamily 1702.01497}}].

\bibitem{Ibe:2016dir}
M.~Ibe, W.~Nakano and M.~Suzuki, \emph{{Constraints on $L_\mu-L_\tau$ gauge
  interactions from rare kaon decay}},
  \href{https://doi.org/10.1103/PhysRevD.95.055022}{\emph{Phys. Rev.}
  {\bfseries D95} (2017) 055022},
  [\href{https://arxiv.org/abs/1611.08460}{{\ttfamily 1611.08460}}].

\bibitem{Alekhin:2015byh}
S.~Alekhin et~al., \emph{{A facility to Search for Hidden Particles at the CERN
  SPS: the SHiP physics case}},
  \href{https://doi.org/10.1088/0034-4885/79/12/124201}{\emph{Rept. Prog.
  Phys.} {\bfseries 79} (2016) 124201},
  [\href{https://arxiv.org/abs/1504.04855}{{\ttfamily 1504.04855}}].

\bibitem{Gninenko:2014pea}
S.~N. Gninenko, N.~V. Krasnikov and V.~A. Matveev, \emph{{Muon g-2 and searches
  for a new leptophobic sub-GeV dark boson in a missing-energy experiment at
  CERN}}, \href{https://doi.org/10.1103/PhysRevD.91.095015}{\emph{Phys. Rev.}
  {\bfseries D91} (2015) 095015},
  [\href{https://arxiv.org/abs/1412.1400}{{\ttfamily 1412.1400}}].

\bibitem{Ge:2017poy}
S.-F. Ge, M.~Lindner and W.~Rodejohann, \emph{{Atmospheric Trident Production
  for Probing New Physics}},
  \href{https://doi.org/10.1016/j.physletb.2017.06.020}{\emph{Phys. Lett.}
  {\bfseries B772} (2017) 164--168},
  [\href{https://arxiv.org/abs/1702.02617}{{\ttfamily 1702.02617}}].

\end{thebibliography}\endgroup

\end{document}